\documentclass[aps,prb,showpacs,twocolumn,longbibliography, amsmath,amssymb,floatfix
]{revtex4-2}

\usepackage{graphicx}
\usepackage{dcolumn}
\usepackage{bm}
\usepackage{mathtools}
\mathtoolsset{showonlyrefs=true}
\usepackage{amsmath}
\usepackage{ragged2e}
\usepackage[usenames,dvipsnames]{color}
\usepackage{siunitx}
\usepackage{accents}
\usepackage[dvipsnames]{xcolor}
\usepackage{booktabs}
\usepackage{braket}
\usepackage{amssymb}
\usepackage{bbm}
\usepackage{hyperref}
\usepackage{mathtools}

\usepackage{placeins}

\begin{document}


\title{Engineering Dirac interface states}

\author{Gabriele Domaine\textsuperscript{1,2}}
\author{Moritz M. Hirschmann\textsuperscript{3}}
\author{Andreas P. Schnyder\textsuperscript{1}}

\affiliation{\textsuperscript{1}Max-Planck-Institut für Festkörperforschung, Heisenbergstrasse 1, D-70569 Stuttgart, Germany}

\affiliation{\textsuperscript{2}Max Planck  Institut f\"ur  Mikrostrukturphysik, Weinberg 2, 06120 Halle, Germany}

\affiliation{\textsuperscript{3}RIKEN Center for Emergent Matter Science, Wako, Saitama 351-0198, Japan}

\date{\today}


\begin{abstract}
We develop a low-energy theory of interface states in anisotropic
multivalley Dirac systems whose masses and kinetic parameters are allowed to vary across
an interface. For sharp interfaces, current-conserving matching conditions
yield analytic expressions for the existence, localization and dispersion of the bound states. We show that the interface velocity is determined by the weighted tangential kinetic terms on the two sides of the seam. Their cancellation can suppress the linear velocity and generate an interface band that is flat to leading order near the projected Dirac point. For the special antisymmetric configuration in which both the Dirac mass and the tangential kinetic coefficient reverse sign with unchanged magnitude, the transparent sharp-interface solution is exactly dispersionless for all conserved momenta within the linear Dirac theory, even though the surrounding bulk bands remain dispersive.
We extend the theory to smooth interfaces, where the modified bound-state envelope generally changes the linear interface velocity through a spatial average of the tangential kinetic coefficient. We also investigate the effects of quadratic corrections in the kinetic \(\sigma_x\) and \(\sigma_y\) channels. To first order in their coefficients and through linear order in the interface momentum, these terms shift the interface-state energy but produce no additional correction to the linear velocity. Finally, we combine continuum and lattice
models to show how interface modes from distinct valleys hybridize
and how the resulting dispersions depend on the microscopic interface properties. Our results establish design principles for controlling the dispersion, localization, and hybridization of Dirac interface states. We further examine two graphene-based mass-domain-wall models as experimentally inspired examples of dispersive copropagating and counterpropagating interface states.
\end{abstract}

\maketitle


\section{Introduction}
\label{sec:introduction}

Interfaces in Dirac materials provide a versatile setting for engineering
one-dimensional electronic modes inside otherwise insulating two-dimensional
systems. Their basic origin can be traced back to the Jackiw--Rebbi mechanism, in
which a sign change of a Dirac mass binds a localized zero mode at a domain
wall~\cite{Jackiw_1976}. A closely related one-dimensional precursor is the
Su--Schrieffer--Heeger model of polyacetylene, where an interface between the
two dimerized ground states supports a localized midgap
state~\cite{Su_1979, Heeger_1988}. In two-dimensional Dirac systems this domain-wall
mechanism acquires an additional momentum direction along the interface, so
that the bound state becomes a dispersing one-dimensional channel~\cite{Callan_1985,Semenoff_2008}.

This mechanism is well established in graphene-based systems. In monolayer graphene, a staggered sublattice potential
opens a mass gap at the Dirac points, and a domain wall across which this mass
changes sign supports propagating midgap modes localized at the domain wall
\cite{Semenoff_2008}. In bilayer graphene, a sign reversal of the interlayer
electric field produces topologically confined interface states~\cite{Martin_2008}, which can be interpreted in terms of valley Chern numbers
\cite{Zhang_2013}. In the standard graphene and bilayer-graphene settings,
where the kinetic chirality of a given valley is fixed across the interface,
a mass inversion is accompanied by a change of the valley-projected
topological index. 

These developments form part of the broader effort to exploit the valley
degree of freedom for device applications~\cite{Schaibley_2016,Vitale_2018}. Within this context,
interface modes have attracted particular interest as controllable one-dimensional transport channels~\cite{Wang_2021, Ren_2016}. Domain-wall intersections have been proposed as
topological current splitters~\cite{Qiao_2014}, while studies of realistic graphene geometries have shown that crystallographic orientation controls valley mixing, and that gate misalignment, finite interface width, and topological defects can modify zero-line-mode dispersion and transport~\cite{Bi_2015}. Valley-polarized interface-state interferometers have been proposed for controlling valley currents and characterizing kink states~\cite{Cheng_2018}, while experiments in bilayer graphene have demonstrated quantized quantum-valley-Hall transport, electrical switching~\cite{Huang_2024}, as well as a filling-factor-dependent redistribution of conduction between domain-wall modes and quantum-Hall edge channels~\cite{Geisenhof_2022}. 


Beyond graphene, general theoretical descriptions of
interface states address different mechanisms under distinct assumptions.
These include mirror-protected states at junctions with opposite Dirac
velocities~\cite{Takahashi_2011}, current-conserving matching at anisotropic
Dirac heterojunctions~\cite{Alspaugh_2024}, tunable chiral edge modes at topological-insulator--magnetic-insulator boundaries~\cite{Beenakker_2024}, and interface states associated
with spin--valley locking~\cite{Zhou_2021} or valley Euler topology~\cite{Ghadimi_2024}.
Recent work on Majorana boundary modes in anisotropic Bogoliubov--de Gennes Dirac systems has also shown that their localization and propagation velocity depend on the velocity tensor and interface orientation~\cite{Paez_2026}.
Two-band theories of conventional semiconductor heterojunctions found
non-topological interface states produced by gap and velocity
mismatch~\cite{Kolesnikov_1998}, with matching conditions later derived from a
tight-binding model~\cite{Kolesnikov_2001}. Smooth topological heterojunctions
can additionally host massive Volkov--Pankratov states~\cite{Tchoumakov_2017},
but this treatment assumes a single isotropic Dirac point with a common
velocity across the interface. In IV--VI heterostructures, valley anisotropy
modifies the interface states and couples them to massive quantum-well
subbands~\cite{Krizman_2022}, but intervalley hybridization is not considered.
Flat or partially flat interface bands have also been predicted in pseudospin-1 \(\alpha-T_3\) systems with symmetry-breaking kink potentials~\cite{Nascimento_2025} and through strain-induced pseudo-Landau quantization~\cite{Tang_2014}, rather than through the controlled cancellation of interface-state velocities. A two-dimensional
model with movable Dirac points produces a valley-mixing gap when the cones
merge~\cite{Denisov_2025}, but the existing work focuses on controlling this
gap rather than the interface-state dispersion and velocity. These works establish several important aspects of Dirac-interface physics, but do not jointly address interfaces with independently varying kinetic and mass parameters on the two sides, the cancellation of their projected tangential contributions, and the subsequent hybridization of modes from multiple valleys. Here we develop such a continuum description and compare its local predictions with full-zone lattice regularizations.

\begin{figure*}[!htb]
\centering
\includegraphics[width=\linewidth]{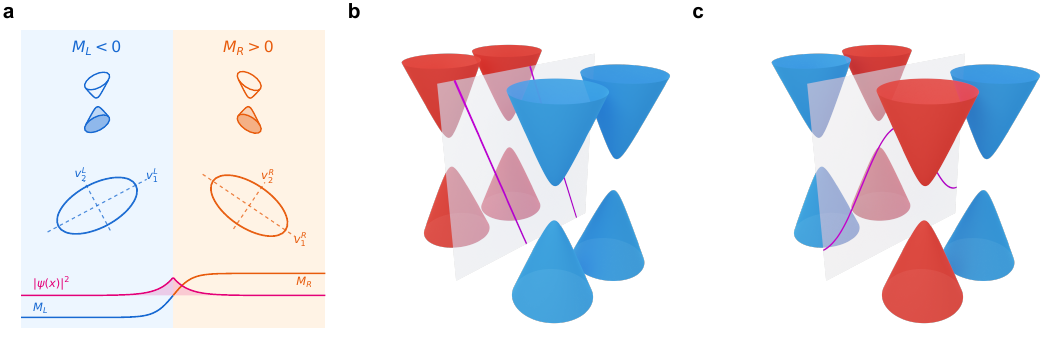}
\caption{\textbf{Copropagating and counterpropagating interface states.}
\textbf{a}) Schematic of an interface between two anisotropic massive Dirac
systems with opposite masses, $M_L<0$ and $M_R>0$. The elliptical
constant-energy contours illustrate the anisotropy and the different
orientations of the principal kinetic axes $v_{1,2}^{L,R}$. The lower profile shows the mass inversion and the
probability density $|\psi(x)|^2$ of the resulting localized interface
state.
\textbf{b}) Two copropagating interface states originating from distinct
Dirac cones.
\textbf{c}) Two counterpropagating interface states. Hybridization between
the modes may open an avoided crossing and connect them into a single
interface band lying inside the bulk gap. The magenta curves in
\textbf{b} and \textbf{c} denote the interface-state dispersions.}
\label{fig:Figure_1}
\end{figure*}

In this work, we develop a unified framework for understanding and controlling interface states in anisotropic multivalley Dirac systems (Fig.~\ref{fig:Figure_1}). We first construct a low-energy theory for anisotropic two-band Dirac systems (Sec.~\ref{subsec:anisotropic_dirac_hamiltonian}), allowing the kinetic terms and masses to differ across a sharp interface. For a sharp interface defined as the zero-width limit of a smooth profile with no singular contact potential, transparent current-conserving matching conditions lead to analytic expressions for the existence of bound states (Sec.~\ref{subsec:bound_states_matching}), as well as for their dispersion (Sec.~\ref{subsec:interface_dispersion}) and for their spatial distribution (Sec.~\ref{subsec:weighted_interface_velocity}). Within this transparent continuum regularization, a mass inversion guarantees the existence of a localized mode, while its velocity depends on how its wave function samples the tangential kinetic terms on the two sides, allowing strong velocity suppression or nearly flat bands through cancellation. When the interface connects cones of opposite kinetic chirality, a mass inversion can bind a mode even when the local valley Chern number is the same on the two sides. For symmetric interfaces with equal kinetic magnitudes, this condition is satisfied when the valley Chern numbers are the same on both sides. We further determine how the interface state energy and dispersion are affected by small Dirac-cone misalignments (Sec.~\ref{subsec:cone_misalignment}), second-order kinetic terms (Sec.~\ref{subsec:quadratic_corrections}), and smooth interface profiles (Sec.~\ref{subsec:smooth_interfaces}).
To address effects beyond the local low-energy description, we study a minimal two-cone lattice model (Sec.~\ref{subsec:lattice_regularized_two_band_model}) and a Wilson-regularized lattice model (Sec.~\ref{subsec:wilson_lattice_model}). The lattice models clarify the role of microscopic matching conditions and reveal how the interface bands behave away from the projected Dirac points, where the local continuum theory is no longer sufficient. In particular, they show how seam hopping, boundary termination, and finite-width profiles affect the dispersion and the spectral separation of the interface states from the bulk. Graphene provides a simple physical setting for illustrating dispersive interface states associated with multiple Dirac valleys. We consider mass domain walls generated by graphene--hexagonal-boron-nitride (hBN) heterostructures and by circularly polarized light (CPL) with opposite helicities (Sec.~\ref{sec:graphene_interfaces}). In this way, the present work extends the established theory of interface modes into a general framework for engineering well isolated anisotropic multivalley Dirac interface states, including nearly dispersionless bands.

\section{Low-energy theory of sharp interfaces}
\label{sec:sharp_interface_theory}

\subsection{Anisotropic Dirac Hamiltonian}
\label{subsec:anisotropic_dirac_hamiltonian}

We start from the operator form of a generic continuum Dirac Hamiltonian in two spatial dimensions 
\begin{gather}
h =
\Gamma_x(x)\,\hat p_x
+
\Gamma_y(x)\,\hat p_y
+
M(x)\,\sigma_3 \,.
\label{eq:H_full_operator_general}
\end{gather}
Throughout the continuum analysis, we use units with \(\hbar = 1\). Here \(h\) acts on a two-component pseudospin degree of freedom, while physical spin is left implicit and consequently each eigenstate is understood to be twofold spin degenerate. The matrices $\sigma_{1,2,3}$ denote the
Pauli matrices associated with a fixed pseudospin frame that is not tied to
the real-space $x$ and $y$ directions, while the matrices $\Gamma_x$ and $\Gamma_y$ are the effective Dirac velocity
matrices associated with motion along the real-space $x$ and $y$ directions
acting in the two-component pseudospin space. The anisotropic Dirac cone may
be parametrized in terms of the principal velocities $v_1,v_2$ and the
orientation angle $\theta$ between the first principal axis and the real-space
$x$ direction. The corresponding momenta along the principal axes are
\begin{equation*}
p_1
=
p_x\cos\theta-p_y\sin\theta,
\qquad
p_2
=
p_x\sin\theta+p_y\cos\theta \,.
\end{equation*}
so that
\begin{align*}
\Gamma_x
&=
v_1\cos\theta\,\sigma_1
+
v_2\sin\theta\,\sigma_2 ,
\\
\Gamma_y
&=
-v_1\sin\theta\,\sigma_1
+
v_2\cos\theta\,\sigma_2 .
\end{align*}
We now introduce a sharp interface centered at $x=0$, across which the local
Dirac parameters
\begin{equation*}
X \in
\left\{
\Gamma_x,\Gamma_y,M
\right\}
\end{equation*}
may change discontinuously
\begin{equation}
X(x)=X^L\Theta(-x)+X^R\Theta(x),
\label{eq:generic_interface_profile}
\end{equation}
where the superscripts $L$ and $R$ denote the values on the left and right
sides of the interface, respectively, and $\Theta(x)$ is the Heaviside step function.
The system remains translationally invariant along $y$, so that
$[h,\hat p_y]=0$ and we may replace $\hat p_y \to k$.
We moreover consider a Dirac crossing located at momentum
$k=K$ along the conserved direction, and define
$q\coloneq k-K$.
On each side of the interface, we perform a local pseudospin rotation (see Appendix~\ref{app:low_energy_hamiltonian}) which aligns the
normal kinetic matrix \(\Gamma_x\) on each side of the interface with the \(\sigma_1\) direction. We denote the Pauli
matrices in this locally aligned pseudospin frame by
\(\sigma_{x,y,z}\). Their subscripts label pseudospin axes rather than
real-space directions. Then, away from the interface, the Hamiltonian on
side $\alpha=L,R$ is given by
\begin{equation}
h_\alpha
=
-i v_\perp^\alpha\sigma_x\partial_x
+
q\left(
u_x^\alpha\sigma_x+u_y^\alpha\sigma_y
\right)
+
M_\alpha\sigma_z \,,
\label{eq:H_side_alpha}
\end{equation}
where
\begin{equation}
v_\perp^\alpha
=
\sqrt{
(v_1^\alpha)^2\cos^2\theta^\alpha
+
(v_2^\alpha)^2\sin^2\theta^\alpha
}
>0
\,,
\end{equation}
is the velocity normal to the interface, while
\begin{align}
u_x^\alpha
&=
\frac{
\left[(v_2^\alpha)^2-(v_1^\alpha)^2\right]
\sin\theta^\alpha\cos\theta^\alpha
}{
v_\perp^\alpha
},
\label{eq:ux_principal_def}
\\
u_y^\alpha
&=
\frac{v_1^\alpha v_2^\alpha}{v_\perp^\alpha}\,,
\label{eq:uy_principal_def}
\end{align}
are, respectively, the components of the tangential kinetic matrix
$\Gamma_y^\alpha$ parallel and perpendicular, in pseudospin space, to the
normal kinetic matrix $\Gamma_x^\alpha$. Thus, both coefficients describe
motion tangential to the interface while their subscripts refer to the axes of the
locally rotated pseudospin frame.

\subsection{Bound states and transparent matching}
\label{subsec:bound_states_matching}

Since we look for bound states localized near $x=0$, a natural choice for the wavefunction is the exponential ansatz $\psi_\alpha(x)
=
A_\alpha e^{\kappa_\alpha x}\chi_\alpha .
\label{eq:ansatz_side}
$
Substituting into $h_\alpha\psi_\alpha=E\psi_\alpha$ gives
the evanescence condition (see Appendix~\ref{app:evanescence_condition})
\begin{equation}
E^2<M_\alpha^2+\left(u_y^\alpha q\right)^2 .
\label{eq:E_condition}
\end{equation}
We define a transparent interface as the zero-width limit of a smooth profile with no singular contact potential. Within this regularization, as shown in Appendix~\ref{app:sharp_interface_details}, this limit generates no additional pseudospin rotation in the locally rotated basis, so that the spinor structure is preserved apart from the factor required by flux normalization. The matching condition is

\begin{equation}
\psi_L(0)
=
\sqrt{\frac{v_\perp^R}{v_\perp^L}}\,
\psi_R(0).
\label{eq:transparent_matching_rotated}
\end{equation}
As shown in Appendix~\ref{app:dispersion_near_q}, at $q=0$ one recovers the Jackiw–Rebbi condition for the existence of the interface state~\cite{Jackiw_1976}
\begin{equation}
\operatorname{sgn}(M_L)
=
-\operatorname{sgn}(M_R)\,,\qquad
E=0\,.
\label{eq:mass_sign_change_condition}
\end{equation}
This mass-inversion criterion applies specifically to the transparent matching condition considered here. For a general current-conserving matching condition, mass inversion is neither necessary nor sufficient for a bound state at \(q=0\) since the interface may prevent the decaying spinors from matching despite a sign change of \(M\), or match them even when \(M\) does not change sign.
\subsection{Interface state dispersion}
\label{subsec:interface_dispersion}

The linear dispersion can be expressed in terms of the valley Chern number $C_{v,\alpha}~
=~
-\frac{1}{2}
\operatorname{sgn}
\left(
u_y^\alpha M_\alpha
\right)$ of a single gapped Dirac point
as
\begin{equation}
E(k)
=
2
\frac{|M_L||M_R|}{|M_L|+|M_R|}
\left[
C_{v,L}\frac{|u_y^L|}{|M_L|}
-
C_{v,R}\frac{|u_y^R|}{|M_R|}
\right]q \,.
\label{eq:general_slope_chern}
\end{equation}
Here \(C_{v,\alpha}\) denotes the valley Chern number obtained by integrating the
Berry curvature of the local massive Dirac cone associated with valley
\(\alpha\). Since this continuum contribution is half-integer, it should not be
confused with the integer Chern number of a complete lattice band. In
particular, equality of the valley Chern numbers on the two sides does not
constitute a mismatch of a global bulk topological invariant.
Equation~\eqref{eq:general_slope_chern} shows that, to first order in \(q\), the group velocity vanishes whenever the two signed ratios
\(C_{v,\alpha}|u_y^\alpha|/|M_\alpha|\) are equal. Since
\(C_{v,\alpha}=\pm 1/2\) and
\(|u_y^\alpha|/|M_\alpha|\geq0\), this cancellation requires
\(C_{v,L}=C_{v,R}\).
A stronger result holds for the special antisymmetric configuration where \(M_R=-M_L\) and \(u_y^R=-u_y^L\).
As shown in Appendix~\ref{app:dispersion_near_q}, in this case the interface mode is exactly dispersionless to all orders in \(q\) throughout the momentum range in which the
linear Dirac description applies. This result does not require equal normal
velocities or any relation between \(u_x^L\) and \(u_x^R\), because these
parameters affect the decay exponents and the spatial phase of the bound state
but drop out of the matching equation for its energy.
In terms of the valley Chern numbers, the antisymmetric configuration gives
\(C_{v,L}=C_{v,R}\). In the more restricted case
\(M_L=-M_R=M\) and \(|u_y^L|=|u_y^R|=|u|\), the linear-order expression
in Eq.~\eqref{eq:general_slope_chern} may equivalently be written as
\begin{equation}
E(k)
=
|u|\,\Delta C_v\,q,
\qquad
\Delta C_v\equiv C_{v,L}-C_{v,R},
\label{eq:symmetric_dispersion_chern}
\end{equation}
which is consistent with the exact result when \(\Delta C_v=0\).

\subsection{Localization and weighted interface velocity}
\label{subsec:weighted_interface_velocity}

Combining these results (see Appendix~\ref{app:Exponential_decay}), the amplitude and decay constant of the ansatz become, to first order in $q$ 
\begin{equation}
A_\alpha
=
\sqrt{
\frac{
2|M_L||M_R|
}{
v_\perp^\alpha\left(|M_L|+|M_R|\right)
}
}\,,\qquad
\kappa_\alpha = \frac{1}{\xi_\alpha}
=
\frac{|M_\alpha|}{|v_\perp^\alpha|}
.
\label{eq:xi_alpha_small_q}
\end{equation}
The amplitude shows how mass asymmetry is reflected in the interface wave
function within the transparent continuum model. Let \(P_L\) and
\(P_R\) denote the normalized probability weights of the interface state on the
left and right sides, for a transparent interface one has (see Appendix~\ref{app:weighted_dispersion_relation})
\begin{align}
P_L
&=
\int_{-\infty}^{0} dx\,
\psi_L^\dagger\psi_L=
\frac{|M_R|}
{|M_L|+|M_R|},
\\
P_R
&=
\int_{0}^{\infty} dx\,
\psi_R^\dagger\psi_R =
\frac{|M_L|}
{|M_L|+|M_R|}.
\label{eq:transparent_weights}
\end{align}
Thus, in this leading transparent-interface description, equal mass magnitudes
give \(P_L=P_R\), whereas unequal mass magnitudes give an asymmetric localized
wave function. More generally, however, equal mass magnitudes do not by
themselves guarantee a symmetric interface state, since microscopic matching
conditions can also modify the wave function. This becomes important for the interface-state dispersion. Within the transparent spinor-preserving matching considered here, the dispersion reduces to first order in \(q\) to
\begin{equation}
E(k)
=
2
\left(
C_{v,L}|u_y^L| P_L
-
C_{v,R}|u_y^R| P_R
\right)q .
\label{eq:weighted_slope_general}
\end{equation}
For the transparent weights in Eq.~\eqref{eq:transparent_weights},
this expression reduces to Eq.~\eqref{eq:general_slope_chern}. 
Thus, Eq.~\eqref{eq:weighted_slope_general} shows that the linear interface velocity vanishes when the weighted tangential contributions satisfy \(C_{v,L}|u_y^L|P_L
=
C_{v,R}|u_y^R|P_R\).
This weighted relation applies more generally to interfaces for which the matching preserves the spinor structure. For a fully general current-conserving matching matrix \(U\), the boundary spinor may also be rotated, and the interface energy and velocity need not be determined by \(P_L\) and \(P_R\) alone.
The special antisymmetric transparent interface discussed above is stronger than this
linear-order cancellation criterion. In that case the full matching equation,
rather than only its expansion near \(q=0\), is solved by
\(E(q)=0\). Exact flatness therefore follows directly from the antisymmetry of
\(M_\alpha\) and \(u_y^\alpha\) and does not rely on separately imposing
\(P_L=P_R\). However, a microscopic lattice interface generally does not satisfy the
same transparent matching condition, so the antisymmetric bulk parameters
alone need not produce an exactly flat lattice band.

\section{Perturbative corrections and finite-width effects}
\label{sec:continuum_extensions}

\subsection{Dirac-cone misalignment and energy offsets}
\label{subsec:cone_misalignment}

Having established the conditions controlling the interface velocity, we now
consider small relative misalignments of the two Dirac cones. Let \(\delta K\)
denote their relative displacement along the conserved momentum direction and
let \(\delta\mu\) denote their relative energy offset. After absorbing the
common momentum and energy shifts, we adopt the symmetric convention in which
the offsets on the left and right sides are, respectively,
\(-\delta K/2\) and \(+\delta K/2\), and
\(-\delta\mu/2\) and \(+\delta\mu/2\). As shown in
Appendix~\ref{app:momentum_and_chemical_shifts}, projection onto the interface
state gives
\begin{equation}
\delta E
=
\left(
C_{v,L}|u_y^L|P_L
+
C_{v,R}|u_y^R|P_R
\right)\delta K
+
\frac{P_L-P_R}{2}\,\delta\mu .
\label{eq:momentum_energy_correction}
\end{equation}
Thus, to first order in \(\delta K\)
and \(\delta\mu\),
the correction is \(q\)-independent and therefore
shifts the interface dispersion without modifying its velocity or the
cancellation condition in Eq.~\eqref{eq:weighted_slope_general}. 

\subsection{Quadratic kinetic corrections}
\label{subsec:quadratic_corrections}

We next examine quadratic-in-momentum corrections to the kinetic Hamiltonian (see Appendix~\ref{app:second_order_kinetic_corrections}). We treat their coefficients perturbatively and retain the projected interface dispersion only through linear order in \(q\). Within this expansion, the quadratic kinetic terms can shift the interface-mode energy at \(q=0\), whereas their first-order correction to the coefficient linear in \(q\) vanishes:
\begin{equation}
E(q)
=
\delta E_0
+
E_1^{(0)}q
+
O(q^2,c^2),
\label{eq:projected_dispersion_first_order}
\end{equation}
where \(E_1^{(0)}\) is the linear coefficient of the unperturbed interface problem, \(\delta E_0\) is the \(q\)-independent first-order correction, and \(c\) collectively denotes the quadratic kinetic coefficients. 
The magnitude of this shift depends on the microscopic structure of the interface, in particular on how the Hamiltonian parameters vary across it. For instance, if the normal velocities differ across the interface, $v_\perp^L\neq v_\perp^R$, the matching condition in Eq.~\eqref{eq:transparent_matching_rotated} produces a discontinuous sharp-interface wave function. The matrix element of the second-order momentum operator $\hat p_x^2$ is then not well defined without resolving the interface profile. Similarly, if the coefficient of the quadratic kinetic term involving only momentum normal to the interface, $c_{xx}^{(y)}$, is assigned a step profile, the resulting interface contribution to $\delta E_0$ is not well defined without an additional microscopic prescription or finite-width regularization. In the special case in which the normal velocity is the same on both sides,
$v_\perp^L=v_\perp^R$, and $c_{xx}^{(y)}$ is constant across the interface,
the zero-width limit of the smooth mass profile in
Eq.~\eqref{eq:smooth_tanh_mass_profile} is finite and gives


\begin{equation}
\delta E_{0} 
=
\operatorname{sgn}(M_R)\,c_{xx}^{(y)}
\left(
\frac{P_L}{\xi_L^2}
+
\frac{P_R}{\xi_R^2}
\right).
\end{equation}
It is also worth noting that these second-order corrections leave the bulk dispersion unchanged
through quadratic order in momentum. Consequently, they provide a means of
tuning the energy separation between the interface mode and the bulk bands.

\subsection{Smooth finite-width interfaces}
\label{subsec:smooth_interfaces}

\begin{figure}[t]
\centering
\includegraphics[width=\columnwidth]{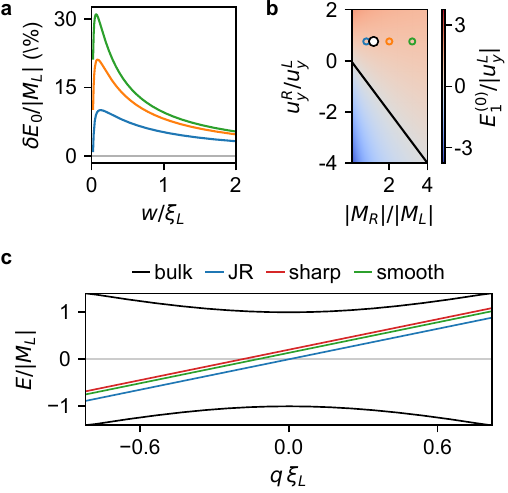}
    \caption{\textbf{Effects of interface smoothing and quadratic kinetic corrections.}
    \textbf{a)} Smooth-interface energy shift
    \(\delta E_0/|M_L|\) as a function of the
    normalized interface width \(w/\xi_L\).
    The blue, orange, and green curves correspond to the parameter sets indicated
    by the small open circles of the same colors in panel (b). These points have
    the same tangential-velocity ratio \(u_y^R/u_y^L\) and different mass ratios
    \(|M_R|/|M_L|\).
    \textbf{b)} Normalized linear interface coefficient
    \(E_1^{(0)}/|u_y^L|\) as a function of the mass ratio
    \(|M_R|/|M_L|\) and tangential-velocity ratio \(u_y^R/u_y^L\).
    The black line denotes the condition \(E_1^{(0)}=0\), for which the
    interface dispersion is flat to first order in \(q\).
    The small colored open circles indicate the parameter sets used in panel (a),
    while the larger open circle marks the parameter set used in panel (c).
    \textbf{c)} Normalized interface-state dispersions and projected bulk edges.
    The black curves show the projected bulk edges, the blue curve the
    Jackiw--Rebbi (JR) dispersion \(E_1^{(0)}q\), and the red and green curves
    the sharp- and smooth-interface results, respectively.
    For the sharp-interface curve, \(c_{xx}^{(y)}\) is spatially constant with
    \(c_{xx}^{(y)}=0.15\). For the smooth-interface calculations,
    \(v_\perp(x)\), \(u_x(x)\), \(u_y(x)\), \(M(x)\), and
    \(c_{xx}^{(y)}(x)\) are interpolated using the same tanh profile, and the
    corresponding smooth zero-mode envelope is used.
    The continuum parameters used in panel (c) are
    \(v_1^L=0.9\), \(v_2^L=1.1\), \(\theta_L=0.1\),
    \(M_L=-1.1\), \(c_{xx}^{(y),L}=0.1\), and
    \(v_1^R=1.1\), \(v_2^R=0.8\), \(\theta_R=0.4\),
    \(M_R=1.3\), \(c_{xx}^{(y),R}=0.2\).
    The smooth-interface width in panel (c) is \(w=0.1\), corresponding to
    \(w\simeq0.12\,\xi_L\simeq0.12\,\xi_R\).}
\label{fig:Figure_4}
\end{figure}

The sharp-interface profile used so far is a useful limiting case because it
reduces the bound-state problem to algebraic matching conditions and gives
closed analytic expressions. However, as discussed above, beyond the linear Dirac approximation the sharp-interface description can lead to ill-defined or regularization-dependent matrix elements when the coefficients of the higher-order terms are discontinuous or when the corresponding sharp-interface wave function is not continuous. Motivated by this, we now consider a smooth finite-width interface in which the spatial variation of the Hamiltonian parameters is resolved. 
As shown in
Appendix~\ref{app:smooth_interface}, for a generic smooth profile, the interface velocity is determined by how the
zero-mode probability density samples \(u_y(x)\).  
A displacement of the bound-state envelope toward one side therefore biases
the velocity toward the local value of \(u_y(x)\), while contributions from
regions with opposite signs of \(u_y(x)\) can cancel.
In particular, for a reflection-antisymmetric smooth interface, this cancellation can again
be exact within the linear Dirac theory.
However, special care is required in the smooth-interface case. A continuous reversal of \(u_y(x)\) necessarily passes through a point where the tangential term \(q\,u_y(x)\sigma_y\) vanishes. At that position the local Hamiltonian has no dispersion linear in the corresponding tangential momentum, so higher-order momentum terms or additional bands may become relevant in a microscopic realization. The exact flatness derived above should therefore be understood as a property of the linear continuum model, whose microscopic validity and momentum range must be checked using a lattice or multiband regularization.
The generic sharp- and smooth-interface dispersions, including the
second-order kinetic corrections discussed above, are compared in
Fig.~\ref{fig:Figure_4}.
The same probability density also makes clear how smoothing affects the
spatial localization. Since the exponential decay is controlled locally by
the ratio \(M(x)/v_\perp(x)\), and a wider interface contains an extended region
where the mass is small compared with its bulk value, the interface state spreads over this region before crossing over to the usual
exponential tails in the two bulks. Thus a smooth interface generally produces a
more spatially extended bound state than the corresponding sharp domain wall.
The same finite-width region can also be viewed as producing a transverse
quantization of eigenvalues. For a narrow
interface, the level spacing associated with the transverse confinement is
large, so the excited levels are pushed into the bulk continuum and are not
resolved as localized subgap modes. However, as the interface is made broader, the
confinement becomes weaker and the corresponding level spacing decreases.
Additional quantized levels can then enter the bulk gap and appear as
massive interface bands. 

\section{Lattice regularizations and microscopic interface effects}
\label{sec:lattice_regularizations}

We now consider how interface states originating from distinct Dirac cones hybridize (see Appendix~\ref{app:two_cone_continuum_model}). As one may expect, these states can hybridize when their projected momenta coincide or when interface scattering supplies the required momentum transfer. Unless the resulting crossing is protected by symmetry, this hybridization opens an avoided-crossing gap.

\subsection{Minimal two-cone lattice model}
\label{subsec:lattice_regularized_two_band_model}

\begin{figure}
    \centering
    \includegraphics[width=\columnwidth]{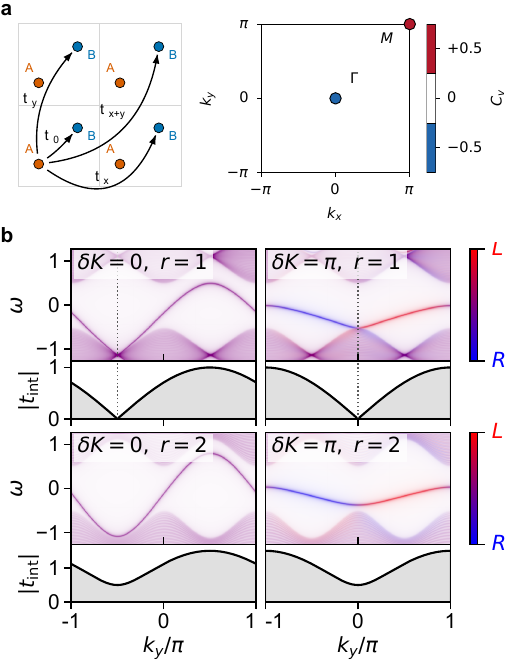}
    \caption{\textbf{Interface-state
    and momentum-dependent seam hopping.}
    \textbf{a}) Left: lattice model showing the four hoppings
    \(t_{\mathbf 0}\), \(t_{\hat x}\), \(t_{\hat y}\), and
    \(t_{\hat x+\hat y}\). Right: Valley Chern number \(C_v\) at the massive
    Dirac valleys of the corresponding gapped bulk lattice model with uniform
    mass term \(m=+0.5\sigma_z\). The markers indicate the discrete valley
    contributions associated with the Dirac points, with opposite signs at the
    two valleys. Reversing the sign of the mass reverses the sign of the local
    valley Chern numbers.
    \textbf{b}) Interface spectral functions and corresponding momentum-dependent seam hopping. The top row shows the isotropic case, while the bottom row shows the anisotropic case; the left and right panels correspond to relative momentum shifts \(\delta K=0\) and \(\delta K=\pi\), respectively. In the isotropic case, \(t_{\mathrm{int}}(k_y)\) vanishes at \(k_y=-\pi/2\) for \(\delta K=0\), where the interface branch touches the bulk band edge. For \(\delta K=\pi\), the zero of the seam hopping shifts to \(k_y=0\), so that the branch associated with the central projected cone touches the bulk bands. In the anisotropic case, by contrast, the seam hopping remains finite for all \(k_y\), with minimum value \(\min |t_{\mathrm{int}}|=1/2\), for both relative momentum shifts. The color scale denotes the edge character of the interface spectral weight: red marks the left edge and blue the right edge, corresponding to the (A) and (B) sublattices under our unit-cell convention. In all panels, the Dirac cones are gapped by a constant mass term \(m=0.5\sigma_z\) with opposite sign across the interface.}
    \label{fig:Figure_5}
\end{figure}

The continuum theory derived above gives controlled local information near a
single projected Dirac cone. It does not, by itself, determine how interface
bands associated with different cones are connected over the full
one-dimensional Brillouin zone. To test the continuum predictions in a
full-zone setting, we therefore introduce a minimal lattice-regularized
two-band toy model. The model is not intended to describe a particular
microscopic material. Its purpose is to provide the simplest periodic
two-band regularization with the minimum number of Dirac cones allowed on a
lattice, namely two. We choose these two cones to be equally separated after
projection onto the interface Brillouin zone. This makes it straightforward
to introduce a relative momentum shift between the two sides of an interface
which exchanges the Dirac cone on one side with the opposite cone on the
other side. The model (see Fig.~\ref{fig:Figure_5}a and Appendix~\ref{app:minimal_lattice_model}) only contains inter-sublattice hopping terms given by
\begin{equation}
\label{eq:lattice_hoppings}
\begin{aligned}
t_{\mathbf 0}
&=
\frac{i}{2},
&
t_{\hat x}
&=
\frac{1}{2},
\\
t_{\hat y}
&=
-\frac{1}{2},
&
t_{\hat x+\hat y}
&=
-\frac{i}{2}.
\end{aligned}
\end{equation} 
For an interface parallel to \(y\), \(k_y\) is
conserved and the model becomes an effective one-dimensional chain along
\(x\). 
When the two sides of the interface are shifted relative to one another in
the conserved momentum direction, the local momentum entering the bulk
Hamiltonian on side \(\alpha=L,R\) is
\begin{equation}
k_\alpha
=
k_y-\delta K_\alpha .
\end{equation}

With this we can define the hopping across the interface seam. We choose a
boundary-adapted unit-cell convention: on each side, the unit-cell origin is
defined at the physical edge site adjacent to the seam, and the semi-infinite
bulk unit cells are then counted away from the interface. Then the interface lies between the last unit cell of
the left domain and the first unit cell of the right domain. The hopping terms
crossing the seam are \(t_{\hat x}\) and \(t_{\hat x+\hat y}\) so that the combined hopping across the interface is
\begin{equation}
t_{\mathrm{int}}(k_y)
=
t_{\hat x}
+
t_{\hat x+\hat y}
e^{
i\left(k_y-\delta K_{\mathrm{int}}\right)
},
\label{eq:tint_general_def}
\end{equation}
where
\begin{equation}
\delta K_{\mathrm{int}}
=
\frac{\delta K_L+\delta K_R}{2}.
\label{eq:deltaKint_midpoint}
\end{equation}
Its magnitude is
\begin{equation}
\left|t_{\mathrm{int}}(k_y)\right|^2
=
\frac{1}{2}
\left[
1+\sin\left(k_y-\delta K_{\mathrm{int}}\right)
\right].
\label{eq:tint_magnitude}
\end{equation}
Thus the seam hopping vanishes whenever
\begin{equation}
k_y
=
\delta K_{\mathrm{int}}
-
\frac{\pi}{2}
\quad
\mathrm{mod}\;2\pi .
\label{eq:tint_zero_condition}
\end{equation}
This condition is independent of the local mass sign. It is instead a
microscopic statement about the destructive interference between the two
lattice hoppings that cross the seam.
For example, when there is no relative momentum shift,
\begin{equation}
\delta K_{\mathrm{int}}=0,
\end{equation}
and Eq.~\eqref{eq:tint_zero_condition} gives
\begin{equation}
t_{\mathrm{int}}(k_y)=0
\qquad
\text{at}
\qquad
k_y=-\frac{\pi}{2}
\quad
\mathrm{mod}\;2\pi .
\label{eq:tint_zero_no_shift}
\end{equation}
When the right side is shifted by \(\pi\),
\begin{equation}
\delta K_{\mathrm{int}}=\frac{\pi}{2},
\end{equation}
the zero is shifted to
\begin{equation}
t_{\mathrm{int}}(k_y)=0
\qquad
\text{at}
\qquad
k_y=0
\quad
\mathrm{mod}\;2\pi \,,
\label{eq:tint_zero_pi_shift}
\end{equation}
as shown in Figs.~\ref{fig:Figure_5}b.
The lattice interface therefore introduces microscopic matching details that are absent from the transparent continuum description. In particular, when
\(\left|t_{\mathrm{int}}(k_y)\right|\) is large, the two sides are strongly
connected across the seam and the interface behaves approximately as a
transparent domain wall. When \(\left|t_{\mathrm{int}}(k_y)\right|\) becomes small, the seam
becomes weakly transmitting and the bound-state problem is no longer described by a transparent continuum matching condition. At a zero of
\(\left|t_{\mathrm{int}}(k_y)\right|\), the seam is effectively cut at that
momentum and the spectrum is controlled instead by the two separate boundary
terminations. For instance, in the present model, these terminations correspond to an
intracell dimerization and do not host isolated edge states. The corresponding
interface band therefore touches or merges with the bulk continuum even
though the local Dirac mass changes sign across the interface. 
A simple way to remove this accidental decoupling at specific $k_y$ is to introduce a kinetic anisotropy parameter \(r\) that rescales the \(k_y\)-dependent part of the massless off-diagonal Hamiltonian relative to the \(k_x\)-dependent part. The isotropic case corresponds to \(r=1\), while \(r\neq1\) produces unequal local Dirac velocities along the two momentum directions. The corresponding interface hopping magnitude is (see Appendix~\ref{app:minimal_lattice_model})
\begin{equation}
\left|t_{\mathrm{int}}(k_y)\right|^2
=
\frac{1}{4}
\left[
r^2+1+2r\sin\left(k_y-\delta K_{\mathrm{int}}\right)
\right],
\label{eq:tint_magnitude_squared_anisotropic}
\end{equation}
with minimum and maximum
\begin{equation*}
\min_{k_y}
\left|
t_{\mathrm{int}}^{(r)}(k_y)
\right|
=
\frac{|r-1|}{2},
\qquad
\max_{k_y}
\left|
t_{\mathrm{int}}^{(r)}(k_y)
\right|
=
\frac{r+1}{2}.
\label{eq:tint_anisotropic_minmax}
\end{equation*}
For \(r=1\), this minimum vanishes and the destructive interference discussed
above is recovered. For \(r\neq1\), the bare seam hopping no longer has an
exact zero. As an example, the spectra for \(r=2\) are shown in the bottom panels of Fig.~\ref{fig:Figure_5}b.
Although the exact zero of the bare seam hopping is removed for \(r=2\), the
separation between the interface band and the bulk continuum can still
remain small. This shows that although a non-zero \(|t_{\mathrm{int}}(k_y)|\) is a necessary condition to have an energy separation and is a useful
diagnostic of microscopic seam transparency, it is not by itself a
complete measure of interface-state localization.
A more general way to formulate this point is to regard the full interface
problem as two terminated half systems coupled by a seam operator
\(V_{\mathrm{int}}(k_y)\). Let
\(\{|\chi_{\lambda,L}(k_y)\rangle\}\) and
\(\{|\chi_{\mu,R}(k_y)\rangle\}\) denote eigenstates of the decoupled left
and right half-space Hamiltonians. These states may include isolated edge
states, when such states exist, as well as bulk or continuum states. The
matrix elements induced by the seam are then
\[
\mathcal T_{\lambda\mu}(k_y)
=
\langle
\chi_{\lambda,L}(k_y)
|
V_{\mathrm{int}}(k_y)
|
\chi_{\mu,R}(k_y)
\rangle .
\]
When both half systems support isolated termination states at the same
conserved momentum, and when these states are well separated from the
corresponding half-space continua, the interface problem can be approximated
by projecting onto this two-dimensional subspace. In that controlled limit the
interface splitting may be interpreted in terms of an effective hopping
between the two termination states. On the other hand, if no isolated termination state exists
for one or both half systems, or if the relevant states are not well separated
from the continuum, this two-state projection is not a good approximation.
There is then no closed two-edge-state subspace and the interface band is
instead controlled by hybridization with the available half-space spectrum,
including bulk-like states, and its separation from the continuum need not be
large even when the bare seam hopping is finite.
This is precisely what happens near the momenta where
\(|t_{\mathrm{int}}(k_y)|\) is smallest. 
The weak separation from the bulk continuum can be understood using the
one-dimensional winding-number criterion for the corresponding decoupled
half-space terminations, as discussed in
Appendix~\ref{app:minimal_lattice_model}. For the unshifted interface,
\(\delta K=0\), the seam hopping is smallest at \(k_y=-\pi/2\). Neither termination
supports an isolated edge state at this momentum. For the shifted interface,
\(\delta K=\pi\), the minimum occurs at \(k_y=0\). This momentum lies at the transition between
localized- and nonlocalized-edge-state regimes and therefore does not support
well-localized isolated edge states. Consequently, removing the exact zero of the seam hopping through anisotropy
does not guarantee an interface band that is well separated in energy from the
projected bulk continuum. This is particularly evident in the bottom left panel in
Fig.~\ref{fig:Figure_5}b, and to a lesser extent in the bottom right panel, since the relevant decoupled terminations do not
provide well-localized edge states that can hybridize into an isolated
interface band.

\subsection{Wilson-regularized lattice model}
\label{subsec:wilson_lattice_model}

\begin{figure}
    \centering
    \includegraphics[width=\columnwidth]{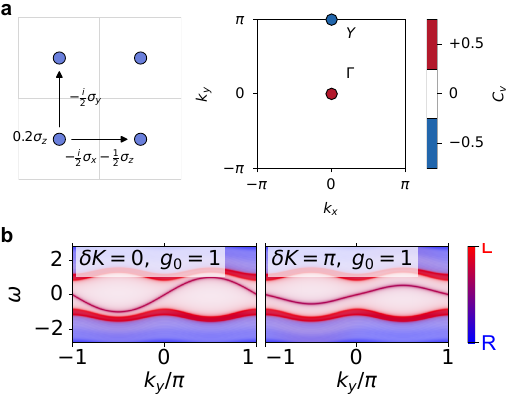}
    \caption{\textbf{Wilson lattice model.}
    \textbf{a}) Real-space hopping structure of the Wilson-regularized lattice model (left), together with the Dirac points in the Brillouin zone (right). The Wilson term gaps out the \(k_x=\pi\) doubler sector, leaving the two Dirac points at \(k_x=0\). \textbf{b}) Interface spectral function for \(\delta K=0\) and \(\delta K=\pi\). The color encodes the side-resolved spectral weight near the interface: red and blue indicate localization predominantly on the left and right sides of the seam, respectively, while purple corresponds to an approximately symmetric distribution. The tendency of the interface band toward red rather than pure purple therefore indicates an asymmetric real-space distribution of the interface state.}
    \label{fig:Figure_7}
\end{figure}

The previous lattice model gives a minimal full-zone regularization with two
projected Dirac cones. It also shows that a finite seam hopping is not enough
to guarantee a well isolated interface band, since at fixed interface momentum,
the decoupled half-space problem does not always contain an isolated edge
state. We therefore introduce a second lattice model in which this limitation
is removed by construction (see Appendix~\ref{app:wilson_edge_state_model}). Specifically, the regularization is chosen such that the one-dimensional winding number normal to the interface is non-zero and independent of the conserved momentum \(k_y\).
Consequently, the decoupled topological half-space supports a localized edge
state for every \(k_y\), providing an isolated state that can hybridize across
the seam throughout the whole interface Brillouin zone.
In real space the onsite block and nearest-neighbor hoppings are (see Fig.~\ref{fig:Figure_7}a)
\begin{gather}
H_0=(m+B)\sigma_z ,
\\
T_x
=
-\frac{i}{2}\sigma_x
-
\frac{B}{2}\sigma_z ,
\qquad
T_y
=
-\frac{i v_y}{2}\sigma_y \,
\end{gather}
respectively.
The hopping across an \(x\)-normal seam is therefore simply \(T_x\). A relative momentum shift between the two sides
is introduced, as before, by shifting the local momentum on one side,
\(k_y\rightarrow k_y-\delta K_\alpha\), or equivalently by multiplying the
corresponding \(y\)-direction hopping by a phase.
The main improvement over the minimal two-cone model is that the existence of
the isolated edge state is independent of \(k_y\) so that the decoupled
topological half-space supports a localized edge state for every conserved
momentum. Thus, as shown in Fig.\ref{fig:Figure_7}b, the interface state never merges with the bulk continuum.

This model also shows how microscopic lattice matching can produce an
interface wave function that is not symmetrically distributed across the
seam. Although the low-energy masses have equal magnitude, the two terminated
lattice systems are not equivalent. The left side is in the edge-state phase,
whereas the right side is trivial for the same termination. After the seam
hopping is restored, the interface state penetrates into both sides, but the
weights \(P_L\) and \(P_R\) need not be equal and the interface velocity can deviate from
the continuum domain-wall solution. This is visible in Fig.~\ref{fig:Figure_7}c,
where the interface band is not purely purple but is shifted toward the red
side of the color scale, indicating a larger spectral weight on one side of the
seam. In the \(\delta K=\pi\) configuration, the interface couples Dirac points
with opposite mass and opposite tangential velocity \(u_y\). For equal left and
right weights, Eq.~\eqref{eq:general_slope_chern} would then predict an exact
cancellation of the two velocity contributions and hence a dispersionless
interface state. This is partially reflected in the smaller group velocity of the interface state compared to Fig.\ref{fig:Figure_7}b. However, because
\(P_L\neq P_R\), the two contributions to the projected velocity do not cancel
completely. 
This microscopic asymmetry can be reduced, for example, by modifying the
interface profile. One possibility is to replace the sharp mass step by a
smooth interpolation over a finite number of lattice spacings. In this case, the
bound state is determined by an extended domain-wall region rather than by a
single termination-dependent matching condition at the seam. For an
approximately antisymmetric mass profile, the envelope samples the two sides of
the interface more evenly, driving the integrated weights \(P_L\) and \(P_R\)
toward equality. Another possibility is to increase the hopping amplitudes on
bonds close to the interface. This locally enhances the coupling between the
two terminated lattices and reduces the tendency of the bound state to remain
concentrated on one side of the seam. 

\begin{figure}
    \centering
    \includegraphics[width=\columnwidth]{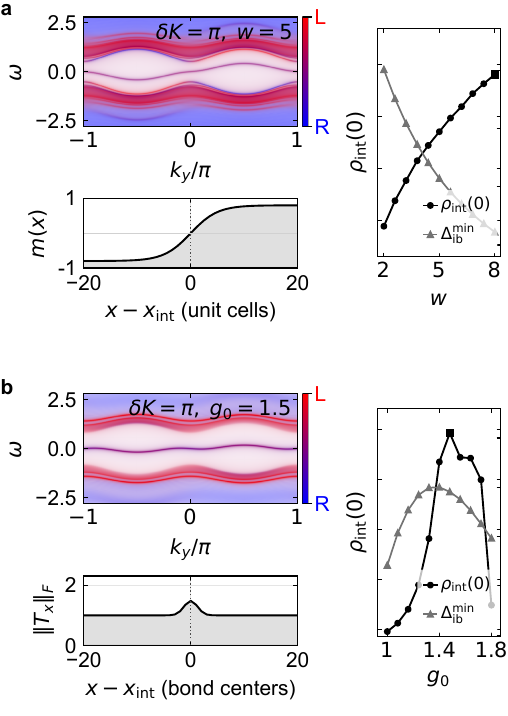}
    \caption{\textbf{Suppression of microscopic left--right asymmetry.}
    \textbf{a}) Smooth-interface. Left: projected spectral function
    for the \(\delta K=\pi\) interface with a smooth mass profile of
    width \(w=5\) as defined in Eq.~\eqref{eq:smooth_tanh_mass_profile} in the Appendix. Red and blue
    denote states localized predominantly on the left and right sides of the
    interface, respectively, while purple indicates comparable weight on both
    sides. Bottom: corresponding mass profile \(m(x)\), interpolating from
    \(m_L=-m_0\) to \(m_R=+m_0\). Right: zero-energy interface-projected
    density of states \(\rho_{\rm int}(0)\), and minimum energy separation of the interface band from the bulk edge or additional localized interface subbands \(\Delta_{\rm ib}^{\min}\).
    \textbf{b}) Interface-hopping tuning for a sharp mass domain wall. Left:
    projected spectral function for enhanced \(x\)-direction hopping near the
    interface, using the peak hopping amplitude \(g_0=1.4\). The hopping \(T_x\) on bond
    centers \(x_b\) is multiplied by
    \(g(x_b)=1+(g_0-1)\exp[-(x_b-x_{\rm int})^2/(2w_{\rm hop}^2)]\),
    with \(w_{\rm hop}=1.25\). Bottom: corresponding Frobenius norm
    \(\|g(x_b)T_x\|_F\) of the local \(x\)-direction hopping. Right:
    \(\rho_{\rm int}(0)\) and \(\Delta_{\rm ib}^{\min}\). The calculation uses \(N_x=96\),
    \(v_y=B=1\), \(m_0=0.8\), \(\delta K=\pi\), and Lorentzian broadening
    \(\eta=0.025\).}
    \label{fig:Figure_8}
\end{figure}

Both mechanisms therefore suppress the
left--right weight imbalance and reduce the residual dispersion of the
\(\delta K=\pi\) interface band, as illustrated in
Fig.~\ref{fig:Figure_8}.
There is, however, an optimal tuning regime rather than a strictly monotonic
improvement. Increasing the smooth-interface width reduces the microscopic
left--right imbalance of the central interface band, but a sufficiently wide
domain-wall region also provides an extended confinement region that can support
additional interface-localized subbands. These subbands enter the spectral gap
and reduce the energy separation between the central interface band and the
rest of the spectrum, as shown in
Fig.~\ref{fig:Figure_8}a. An analogous compromise occurs when the hopping across
a sharp interface is enhanced. Moderate increases of the interface hopping
improve hybridization across the seam, whereas too large a hopping makes the
interface band more dispersive and less isolated in energy. The useful regime
is therefore again an intermediate range, as shown in Fig.~\ref{fig:Figure_8}b.

\section{Graphene mass-domain-wall realizations}
\label{sec:graphene_interfaces}

Graphene provides a minimal material setting in which dispersive mass-domain-wall interface states can occur.
At low energies, the two gapped valleys of graphene are described by
massive Dirac Hamiltonians of the form~\cite{CastroNeto_2009}
\begin{gather}
h_\nu(\mathbf q,x)
=
v_F
\left(
\nu q_x\sigma_x
+
q_y\sigma_y
\right)
+
M_\nu(x)\sigma_z ,
\\
\nu=\pm1 ,
\end{gather}
where \(\nu=+1\) and \(\nu=-1\) label the \(K\) and \(K'\) valleys,
respectively, \(v_F\) is the Fermi
velocity, and the Pauli matrices \(\sigma_{x,y,z}\) act in the sublattice space. The Hamiltonian is written for one spin sector, so in the absence of spin-dependent interactions, all modes acquire an additional twofold physical-spin degeneracy. The valley-dependent mass may be decomposed as
\[
M_\nu(x)
=
m_S(x)
+
\nu m_H(x).
\]
Here \(m_S\) is a sublattice-staggered Semenoff mass, which breaks inversion
symmetry \cite{Semenoff_1984}. Such a mass can be induced by a
sublattice-asymmetric environment such as hexagonal boron nitride (hBN)
\cite{Giovannetti_2007,Jung_2015}. The second contribution, \(m_H\), is a
Haldane mass, which breaks time-reversal symmetry and realizes a
Chern-insulating gap~\cite{Haldane_1988}. In graphene, an effective Haldane mass can be
generated dynamically by circularly polarized light (CPL)~\cite{Oka_2009,Kitagawa_2011,McIver_2019}.

\subsection{Haldane and Semenoff mass domain walls}
\label{subsec:graphene_mass_domain_walls}

As a concrete example, we focus on a zigzag interface. In this orientation,
the two valley projections are distinct in the interface Brillouin zone, so that
intervalley mixing is small.
We first consider the case in which the Haldane mass changes sign across the
interface, while the Semenoff mass is absent. The valley masses then satisfy
\begin{equation}
M_\nu^L=\nu m_H^L,
\qquad
M_\nu^R=\nu m_H^R,
\qquad
m_H^L m_H^R<0 .
\end{equation}
Thus the mass is inverted in both valleys. However, because the Haldane mass
itself changes sign between valleys, this mass pattern compensates the
opposite kinetic chirality of the two graphene Dirac cones. The resulting interface modes therefore propagate with the same chirality,
forming a copropagating pair.

We next consider the case in which the Semenoff mass changes sign across the
interface, while the Haldane mass is absent. The valley masses are then
\[
M_\nu^L=m_S^L,
\qquad
M_\nu^R=m_S^R,
\qquad
m_S^L m_S^R<0 .
\]
Again, the mass is inverted in both valleys, so each valley separately
supports an interface mode. In contrast to the Haldane case, however, the mass
now has the same sign in the two valleys. Since the two graphene cones have opposite kinetic chirality, the
corresponding interface modes propagate in opposite directions. The complete
low-energy spinless spectrum therefore contains a counterpropagating pair. These copropagating and counterpropagating structures
are illustrated by the full-Brillouin-zone calculation in the upper part of
Fig.~\ref{fig:Figure_2}a, where an arbitrarily large mass and a sharp interface are
used simply to illustrate the relative chiralities of the modes at the two
valleys.
On the other hand, the low-energy spectra in the lower part of
Fig.~\ref{fig:Figure_2}b-c, which sample a small momentum window centered on
\(K\), use a realistic mass magnitude and a smooth interface profile as explained below. 
For the Semenoff mass interface, we consider a graphene monolayer on top of an interface formed by an hBN
inversion-domain boundary across which boron and nitrogen interchange between
the two inequivalent honeycomb sublattices~\cite{Taha_2017,Elder_2023}. For a fixed graphene
crystallographic orientation, this interchange reverses the
sublattice-dependent substrate potential and therefore the sign of the
effective Semenoff mass. 
Structural relaxation can
nevertheless accompany such a boundary. In particular, calculations for a
graphene/hBN bilayer predict that graphene spanning an hBN inversion boundary
develops a localized displacement wall, together with in-plane strain and
out-of-plane corrugation \cite{Elder_2023}.
However, the quantitative spatial extent of this relaxation was found to depend
on the microscopic setup. We therefore chose \(w_S=7.5\,\mathrm{nm}\) in
Fig.~\ref{fig:Figure_2}c as a phenomenological estimate for the smoothing length
of the induced Semenoff-mass profile. Our main conclusions concerning the
counterpropagating interface modes do not rely on this particular choice of
\(w_S\).

Grain boundaries have been observed experimentally in CVD-grown hBN
\cite{Gibb_2013,Li_2015}. However, the cited measurements did not directly
resolve the boron and nitrogen assignment at the boundaries and therefore did
not establish the extended inversion-domain boundary considered here. A
controlled graphene device spanning such a zigzag hBN inversion boundary has
also not, to our knowledge, been characterized spectroscopically. We therefore
regard this structure as a physically motivated realization of the effective
mass-domain-wall model rather than as an experimentally established device
geometry.

For the CPL-induced Haldane mass, the sign of the Floquet mass is controlled by
the helicity of the light \cite{Oka_2009,Kitagawa_2011,McIver_2019}, such that an
interface between regions of opposite helicity can realize the
required mass inversion. A closely related geometry consisting of two
counter-rotating beams has been proposed in three-dimensional topological insulators
\cite{Calvo_2015}. The spatial variation of the induced mass is set by the
optical field profile. For freely propagating mid-infrared light, such spatial
variation would occur on optical wavelength scales, whereas a nanoscale
interface would require near-field or plasmonic confinement. Mid-infrared
graphene plasmons have been experimentally realized in nanoresonators with
dimensions down to approximately \(15\,\mathrm{nm}\) \cite{Brar_2013}, and
plasmonic nanoantenna calculations show that circularly polarized near fields
can be localized on tens-of-nanometers scales \cite{Ogut_2009}. Although these studies do not explicitly model an interface between regions of
opposite helicity, they establish nanoscale length scales over which plasmonic
near fields can vary. We therefore take \(w_H=15\,\mathrm{nm}\) in
Fig.~\ref{fig:Figure_2}b as a plausible scale over which the optical helicity,
and hence the CPL-induced Haldane mass, could change sign. Importantly, the chirality of the
resulting interface modes is determined solely by the sign reversal of the Haldane
mass and does not depend on the precise value of the interface width.

We model both interfaces using the smooth interpolation
\begin{equation}
m_\lambda(x)
=
\frac{m_\lambda^L+m_\lambda^R}{2}
+
\frac{m_\lambda^R-m_\lambda^L}{2}
\tanh\frac{x}{w_\lambda},
\qquad
\lambda=S,H .
\label{eq:graphene_smooth_tanh_mass_profile}
\end{equation}
In both cases we choose a bulk gap of
\(\Delta=20\,\mathrm{meV}\). For graphene on hBN, this value is consistent
with theoretical estimates for aligned graphene/hBN including structural
relaxation and many-body effects \cite{Jung_2015}. For the CPL realization,
it lies within the experimentally relevant tens-of-meV scale of
light-induced Floquet gaps in graphene \cite{McIver_2019}. We use the same
gap in the two realizations to facilitate their direct comparison.
This gap corresponds to a Semenoff mass
\(\lvert m_S\rvert=\Delta/2=10\,\mathrm{meV}\) for graphene on hBN, and a purely imaginary next-nearest-neighbor hopping~\cite{Haldane_1988}
\(t_2=\Delta/(6\sqrt{3})\simeq1.92\,\mathrm{meV}\) for the CPL model.
We use the standard graphene nearest-neighbor hopping
\(t=2.7\,\mathrm{eV}\) and lattice constant
\(a=0.246\,\mathrm{nm}\), consistent with conventional tight-binding
parameters for graphene~\cite{Reich_2002,Cheng_2017}.

The upper part of Fig.~\ref{fig:Figure_2}a presents an
illustrative full-Brillouin-zone lattice calculation for zigzag
ribbons with sharp Haldane- and Semenoff-mass domain walls. An intentionally
large mass \(m/t=0.5\) is used solely to make the interface branches clearly
visible. The calculation is not intended to represent realistic energy or
length scales, but directly illustrates the copropagating CPL modes and the
counterpropagating hBN modes. The upper parts of
Figs.~\ref{fig:Figure_2}b and \ref{fig:Figure_2}c illustrate,
respectively, the opposite-helicity CPL domains and graphene above an hBN
inversion-domain boundary. The lower part of Fig.~\ref{fig:Figure_2}a
compares the transparent sharp-interface result with the smooth-interface
continuum predictions obtained using the same corresponding widths for the CPL and hBN cases. For the symmetric
mass inversions considered here, the Fermi velocity is spatially uniform.
Consequently, smoothing modifies the transverse envelope of the bound state
but not its leading Dirac dispersion, and the three analytical curves nearly
coincide. The lower parts of Figs.~\ref{fig:Figure_2}b and
\ref{fig:Figure_2}c show, respectively, the interface-projected lattice spectral functions for the CPL and hBN domain
walls in a momentum window around \(K\) for a transverse ribbon width of approximately \(374\,\mathrm{nm}\). Near this projected Dirac point, the
principal interface branch follows the low-energy prediction, while the
surrounding ribbon subbands form the projected bulk continuum. 

{\color{blue}
\begin{figure*}[t]
    \centering
    \includegraphics[width=\textwidth]{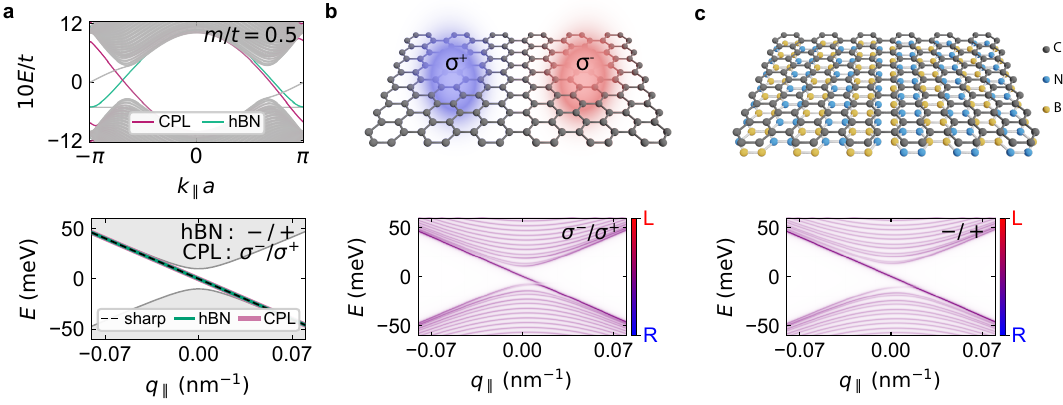}
    \caption{\textbf{Graphene mass-domain-wall interfaces.}
    Top row: \textbf{(a)} Illustrative full-Brillouin-zone spectra for sharp
    Haldane- and Semenoff-mass domain walls. The intentionally large mass
    \(m/t=0.5\) is used only to make the mode chiralities visible: the CPL modes
    are copropagating, whereas the hBN modes are counterpropagating.
    \textbf{(b)} Graphene regions driven by opposite CPL helicities.
    \textbf{(c)} Graphene above an hBN inversion-domain boundary.
    Bottom row: \textbf{(a)} Continuum dispersion near the projected \(K\) point
    for a transparent sharp interface and for smooth hBN and CPL profiles with
    \(w_S=7.5\,\mathrm{nm}\), \(w_H=15\,\mathrm{nm}\), and
    \(\Delta=20\,\mathrm{meV}\). The gray regions denote the projected bulk
    continuum. \textbf{(b)} Interface-projected spectral function for the smooth
    CPL-induced Haldane-mass domain wall near \(K\). \textbf{(c)} Corresponding
    spectral function for the smooth hBN-motivated Semenoff-mass domain wall.
    In the bottom row, \(q_\parallel=k_\parallel-K_\parallel\), with
    \(K_\parallel=+2\pi/(3a)\). Red, blue, and purple spectral weight indicate
    predominantly left-localized, right-localized, and equally distributed
    states, respectively.}
    \label{fig:Figure_2}
\end{figure*}
}

\section{Summary and outlook}
\label{sec:conclusion}

We have developed a unified low-energy framework for interface states in anisotropic multivalley Dirac systems with spatially varying masses and kinetic parameters. For transparent sharp interfaces, we derived the conditions for the existence and localization of bound states and obtained their dispersion in Sec.~\ref{sec:sharp_interface_theory}. In particular, the interface velocity is determined by the weighted tangential kinetic contributions from the two sides of the interface (Secs.~\ref{subsec:interface_dispersion} and \ref{subsec:weighted_interface_velocity}). These contributions may reinforce or cancel, providing a general mechanism for suppressing the interface velocity and generating nearly dispersionless bands without requiring the surrounding bulk bands to be flat. For the special antisymmetric configuration in which both the Dirac mass and the tangential kinetic coefficient reverse sign, this cancellation becomes exact within the linear Dirac theory.
We investigated the robustness of this mechanism against several corrections in Sec.~\ref{sec:continuum_extensions}. Small relative momentum and energy offsets of the Dirac cones shift the interface-state energy without changing its linear velocity (Sec.~\ref{subsec:cone_misalignment}), while quadratic kinetic corrections produce an energy shift but no additional first-order correction to the velocity (Sec.~\ref{subsec:quadratic_corrections}). For smooth interfaces, the velocity is instead controlled by a spatial average of the tangential kinetic coefficient over the interface-state probability density (Sec.~\ref{subsec:smooth_interfaces}). The lattice regularizations of Sec.~\ref{sec:lattice_regularizations} further show how these continuum results are modified by microscopic seam matching, boundary termination, intervalley hybridization, and the full band structure away from the projected Dirac points. In particular, they demonstrate that interface smoothing and local modification of the seam coupling can be used to control both the residual dispersion and the spectral isolation of the interface band. Finally, in Sec.~\ref{sec:graphene_interfaces} we illustrated the framework using graphene mass-domain walls: opposite-helicity circularly polarized light generates copropagating interface modes, whereas a reversal of the sublattice-staggered mass, as motivated by a graphene--hBN inversion-domain boundary, generates counterpropagating modes.

The framework is not restricted to these graphene realizations. More generally, it can be applied to interfaces between gapped Dirac valleys whenever the mass, kinetic anisotropy, or orientation of the Dirac cones can be controlled spatially. Possible electronic settings include electrically gated bilayer graphene \cite{Li_2016}, AB--BA stacking domain walls in bilayer graphene \cite{Yin_2016}, and magnetic domain boundaries on topological-insulator surfaces \cite{Rusinov_2021}, in addition to optically engineered interfaces such as counter-rotating illumination domains \cite{Calvo_2015}.

Future theoretical work could extend the present framework beyond the transparent matching convention to more general current-conserving interface conditions and establish their relation to microscopic seam structure. Further extensions could address disorder and intervalley scattering, and interaction effects in strongly velocity-suppressed interface bands.

In an experimental or device setting, local strain could provide a possible means of modifying the hopping amplitudes near the interface and thereby tuning the interface-state dispersion, localization, and spectral isolation. This possibility may be particularly relevant for graphene--hBN heterostructures, where lattice mismatch and structural relaxation can produce spatially nonuniform strain, including near domain-wall regions \cite{Woods_2014}. More broadly, the ability to control the velocity, propagation direction, and localization of interface modes could enable tunable one-dimensional transport channels for valleytronic applications, while strongly velocity-suppressed or nearly flat interface bands may provide a platform for enhanced interaction effects and correlated one-dimensional phases.

\section{Acknowledgments}

G.D. acknowledges support by the Max Planck Graduate Center for Quantum Materials (MPGC-QM). M.M.H. is funded by the Deutsche Forschungsgemeinschaft (DFG, German Research Foundation) - project number 518238332 and by the RIKEN Special Postdoctoral Researcher Program. A.P.S. is funded by the Deutsche Forschungsgemeinschaft (DFG, German Research Foundation) – TRR 360 – 492547816. 

\appendix

\section{Effective Low-Energy Hamiltonian}
\label{app:low_energy_hamiltonian}

Since the product $\Gamma_x(x)\hat p_x$ is not Hermitian, we
choose a symmetric operator ordering~\cite{Linnik_2012, de_Juan_2012}, so that the kinetic term in
the $x$ direction becomes
\begin{equation}
\begin{aligned}
\Gamma_x(x)\hat p_x
&\longrightarrow
\frac{1}{2}\{\Gamma_x(x),\hat p_x\}
\\
&=
-i\left(
\Gamma_x(x)\partial_x
+\frac{1}{2}\,\partial_x\Gamma_x(x)
\right).
\end{aligned}
\label{eq:sym_vxpx}
\end{equation}

Thus, away from the interface, the Hamiltonian on side $\alpha=L,R$ becomes
\begin{equation}
h_\alpha
=
-i\Gamma_x^\alpha \partial_x
+
\Gamma_y^\alpha q
+
M_\alpha\sigma_3 .
\label{eq:H_side_alpha_global}
\end{equation}

On each side of the interface, we perform a local pseudospin gauge rotation. Here, $\sigma_{1,2,3}$ refer to the original fixed pseudospin frame, whereas
$\sigma_{x,y,z}$ below refer to the locally rotated frame, whose $x$ axis is
chosen to be parallel to the normal kinetic matrix. We define
\begin{equation}
S_\alpha
=
e^{-i\varphi_\alpha\sigma_3/2},
\qquad
\varphi_\alpha
=
\operatorname{atan2}
\left(
v_2^\alpha\sin\theta^\alpha,
v_1^\alpha\cos\theta^\alpha
\right),
\label{eq:S_alpha_def}
\end{equation}
where $\operatorname{atan2}(y,x)$ denotes the polar angle of the vector
$(x,y)$. Thus, in the rotated basis on side $\alpha$ one may write
\begin{equation}
S_\alpha^\dagger \Gamma_x^\alpha S_\alpha
=
v_\perp^\alpha\sigma_x,
\qquad
S_\alpha^\dagger \Gamma_y^\alpha S_\alpha
=
u_x^\alpha\sigma_x+u_y^\alpha\sigma_y .
\label{eq:rotated_gamma_def}
\end{equation}
Here
\begin{equation}
v_\perp^\alpha
=
\sqrt{
(v_1^\alpha)^2\cos^2\theta^\alpha
+
(v_2^\alpha)^2\sin^2\theta^\alpha
}
>0
\,,
\end{equation}
is the velocity normal to the interface while
\begin{align}
u_x^\alpha
&=
\frac{
\left[(v_2^\alpha)^2-(v_1^\alpha)^2\right]
\sin\theta^\alpha\cos\theta^\alpha
}{
v_\perp^\alpha
},
\label{eq:ux_principal_def_appendix}
\\
u_y^\alpha
&=
\frac{v_1^\alpha v_2^\alpha}{v_\perp^\alpha}\,,
\label{eq:uy_principal_def_appendix}
\end{align}
are respectively the component of the tangential kinetic matrix parallel and
perpendicular to the normal kinetic matrix.

In the locally rotated basis, the Hamiltonian on side $\alpha=L,R$ is
\begin{equation}
h_\alpha
=
-i v_\perp^\alpha\sigma_x\partial_x
+
q\left(
u_x^\alpha\sigma_x+u_y^\alpha\sigma_y
\right)
+
M_\alpha\sigma_z .
\label{eq:H_side_alpha_appendix}
\end{equation}

\section{Evanescence condition}
\label{app:evanescence_condition}

We look for bound states localized near $x=0$. On each side we therefore use
the exponential ansatz
\begin{equation}
\psi_\alpha(x)
=
A_\alpha e^{\kappa_\alpha x}\chi_\alpha .
\label{eq:ansatz_side_appendix}
\end{equation}

Substituting into $h_\alpha\psi_\alpha=E\psi_\alpha$ gives
\begin{equation}
A_\alpha \left[
\left(u_x^\alpha q-i v_\perp^\alpha\kappa_\alpha\right)\sigma_x
+
u_y^\alpha q\,\sigma_y
+
M_\alpha\sigma_z
-
E\sigma_0
\right]\chi_\alpha=0 .
\label{eq:bulk_spinor_eq}
\end{equation}

The determinant condition gives
\begin{equation}
\kappa_\alpha
=
\frac{
-i u_x^\alpha q
\pm
\sqrt{
M_\alpha^2+\left(u_y^\alpha q\right)^2-E^2
}
}
{v_\perp^\alpha}.
\label{eq:kappa_alpha_general}
\end{equation}

For a localized interface mode,
\begin{equation*}
\Re(\kappa_L)>0,
\qquad
\Re(\kappa_R)<0 .
\end{equation*}

Hence the decaying solutions are
\begin{align}
\kappa_L
&=
\frac{
-i u_x^L q
+
\sqrt{
M_L^2+\left(u_y^L q\right)^2-E^2
}
}
{v_\perp^L},
\label{eq:kappa_L_decay}
\\
\kappa_R
&=
\frac{
-i u_x^R q
-
\sqrt{
M_R^2+\left(u_y^R q\right)^2-E^2
}
}
{v_\perp^R}.
\label{eq:kappa_R_decay}
\end{align}

The evanescence condition is therefore
\begin{equation}
E^2<M_\alpha^2+\left(u_y^\alpha q\right)^2 .
\label{eq:E_condition_appendix}
\end{equation}

\section{Transparent interface}
\label{app:sharp_interface_details}

In general, for a sharp interface one may impose a linear matching condition
of the form
\begin{equation}
\Psi_L(0)=T\,\Psi_R(0),
\label{eq:transparent_matching_global}
\end{equation}
where $\Psi_\alpha=S_\alpha\psi_\alpha$ denotes the spinor in the original
pseudospin basis, while $\psi_\alpha$ denotes the spinor in the locally
rotated basis used above. Therefore the matching condition for the rotated
spinors is
\begin{equation}
\psi_L(0)=S_L^\dagger T S_R\,\psi_R(0).
\label{eq:transparent_matching}
\end{equation}
The matrix \(T\) must preserve the normal Dirac current,
\begin{equation*}
j_x^L=j_x^R,
\qquad
j_x^\alpha
=
\Psi_\alpha^\dagger\Gamma_x^\alpha\Psi_\alpha
=
\psi_\alpha^\dagger v_\perp^\alpha \sigma_x \psi_\alpha .
\end{equation*}

Current conservation for arbitrary boundary spinors therefore requires
\begin{equation}
T^\dagger\Gamma_x^L T
=
\Gamma_x^R .
\label{eq:current_conservation_global}
\end{equation}
Equivalently, in the locally rotated bases,
\begin{equation}
(S_L^\dagger T S_R)^\dagger
\sigma_x
(S_L^\dagger T S_R)
=
\frac{v_\perp^R}{v_\perp^L}
\sigma_x .
\label{eq:STR_sigma}
\end{equation}

Since \(v_\perp^R,v_\perp^L>0\), we can define
\begin{equation}
S_L^\dagger T S_R
=
\sqrt{\frac{v_\perp^R}{v_\perp^L}}\,U,
\end{equation}
so that from Eq.~\eqref{eq:transparent_matching}
\begin{equation}
\psi_L(0)=\sqrt{\frac{v_\perp^R}{v_\perp^L}}\,U\,\psi_R(0),
\label{eq:transparent_matching_U}
\end{equation}
and from Eq.~\eqref{eq:STR_sigma}
\begin{equation}
U^\dagger\sigma_xU=\sigma_x .
\label{eq:current_conserving_U}
\end{equation}
Equation~\eqref{eq:current_conserving_U} characterizes the family of
current-conserving sharp-interface boundary conditions, but does not by itself
determine \(U\). To select a particular member of this family, we define the
transparent interface as the zero-width limit of a smooth profile
\begin{equation*}
\Gamma_x(x)
=
v_\perp(x)S(x)\sigma_xS^\dagger(x),
\qquad
S(x)=e^{-i\varphi(x)\sigma_3/2},
\end{equation*}
with no additional contact potential or other singular interface term. Writing
\(\Psi(x)=S(x)\psi(x)\) and
\begin{equation*}
A(x)
=
S^\dagger(x)\partial_xS(x)
=
-\frac{i}{2}(\partial_x\varphi)\sigma_z,
\end{equation*}
the normal kinetic operator in Eq.~\eqref{eq:sym_vxpx}, transformed to the
locally rotated basis, acts on \(\psi\) as
\begin{align}
\frac{1}{2}
S^\dagger\{\Gamma_x(x),\hat p_x\}S\psi
&=
-i\Bigl[
v_\perp\sigma_x\partial_x\psi
+
\frac{1}{2}(\partial_xv_\perp)\sigma_x\psi
\nonumber\\
&\hspace{1.2cm}
+
\frac{v_\perp}{2}\{\sigma_x,A\}\psi
\Bigr]
\nonumber\\
&=
-i\sqrt{v_\perp}\,\sigma_x
\partial_x\left(\sqrt{v_\perp}\,\psi\right).
\label{eq:rotated_normal_kinetic_smooth}
\end{align}
For the last equality we used
\(\{\sigma_x,A\}=0\). Thus, the spatial variation of the local pseudospin
rotation produces no additional contact rotation.
Combining Eq.~\eqref{eq:rotated_normal_kinetic_smooth} with the mass and
tangential kinetic terms, the full Schr\"odinger equation within the smooth
interface region is
\begin{align}
-i\sqrt{v_\perp}\,\sigma_x
\partial_x\left(\sqrt{v_\perp}\,\psi\right)
+
\left[
q\left(u_x\sigma_x+u_y\sigma_y\right)
+
M\sigma_z
\right]\psi
=
E\psi .
\label{eq:smooth_interface_schrodinger_rotated}
\end{align}
After rearranging, this becomes
\begin{align}
\partial_x\left(\sqrt{v_\perp}\,\psi\right)
=
\frac{i}{\sqrt{v_\perp}}\,
\sigma_x
\left[
E
-
q\left(u_x\sigma_x+u_y\sigma_y\right)
-
M\sigma_z
\right]\psi .
\label{eq:flux_spinor_derivative}
\end{align}
We assume that \(v_\perp(x)\) remains non-zero and that
\(u_x(x)\), \(u_y(x)\), and \(M(x)\) do not diverge as
\(w\rightarrow0\). For fixed finite \(E\) and \(q\), and a finite interface
spinor, the right-hand side is then bounded from above and its integral over the interface width vanishes in the zero-width
limit. Using
\(\lim_{w\to0}v_\perp(-w/2)=v_\perp^L\) and \(\lim_{w\to0}v_\perp(w/2)=v_\perp^R\)
we obtain
\begin{equation}
\sqrt{v_\perp^R}\,\psi_R(0) -\sqrt{v_\perp^L}\,\psi_L(0)
=
0.
\label{eq:flux_normalized_continuity}
\end{equation}
Equivalently,
\begin{equation}
\psi_L(0)
=
\sqrt{\frac{v_\perp^R}{v_\perp^L}}\,
\psi_R(0).
\label{eq:transparent_matching_rotated_appendix}
\end{equation}
Comparison with Eq.~\eqref{eq:transparent_matching_U} therefore
selects \(U=\sigma_0\) up to an irrelevant overall phase.
Consequently, the matching matrix in the original pseudospin basis is
\begin{equation}
T
=
\sqrt{\frac{v_\perp^R}{v_\perp^L}}\,
S_LS_R^\dagger .
\label{eq:T_transparent_definition}
\end{equation}

Thus, \emph{transparent} here denotes the absence of an additional point-like interface
interaction. Modified seam hoppings, interface reconstruction,
or contact scattering would in general correspond to a nontrivial \(U\)
satisfying Eq.~\eqref{eq:current_conserving_U}.

\section{Sharp-Interface Dispersion}
\label{app:dispersion_near_q}

A convenient choice of eigenspinor in Eq.~\eqref{eq:ansatz_side_appendix} is
\begin{equation}
\chi_\alpha
=
\begin{pmatrix}
\left(u_x^\alpha-i u_y^\alpha\right)q
-i v_\perp^\alpha\kappa_\alpha\\[3pt]
E-M_\alpha
\end{pmatrix}.
\label{eq:chi_alpha_explicit}
\end{equation}

Then, Eq.~\eqref{eq:transparent_matching_rotated_appendix}
may be written as
\begin{equation}
\frac{M_L-E}{D_L}
=
\frac{M_R-E}{D_R},
\label{eq:matching_single_equation}
\end{equation}
where
\begin{align}
D_L
&=
\sqrt{M_L^2+\left(u_y^Lq\right)^2-E^2}
+
u_y^Lq,
\\
D_R
&=
-\sqrt{M_R^2+\left(u_y^Rq\right)^2-E^2}
+
u_y^Rq .
\label{eq:DL_DR_def}
\end{align}

We first consider the special antisymmetric configuration
\begin{equation}
M_R=-M_L,
\qquad
u_y^R=-u_y^L .
\label{eq:exact_flat_antisymmetric_condition_appendix}
\end{equation}
Under this condition, the two square roots in
Eq.~\eqref{eq:DL_DR_def} are identical for arbitrary \(E\), so that \(D_R=-D_L\).
Then, using Eqs.~\eqref{eq:exact_flat_antisymmetric_condition_appendix}, the matching condition
in Eq.~\eqref{eq:matching_single_equation} becomes
\begin{equation}
\frac{M_L-E}{D_L}
=
\frac{-M_L-E}{-D_L}
=
\frac{M_L+E}{D_L}.
\end{equation}
Since \(D_L\neq0\) for an evanescent solution with nonzero asymptotic mass,
this equation requires \(E=-E\)
and therefore
\begin{equation}
E(q)=0 .
\label{eq:exact_flat_dispersion_appendix}
\end{equation}
No expansion in \(q\) has been made, so the
cancellation holds to all orders in \(q\) within the linear Dirac
Hamiltonian. Notice that \(v_\perp^\alpha\) and \(u_x^\alpha\) do not enter
Eq.~\eqref{eq:matching_single_equation}; they modify the decay exponents in
Eqs.~\eqref{eq:kappa_L_decay} and \eqref{eq:kappa_R_decay}, but not the
energy of the interface state.
The evanescence condition in Eq.~\eqref{eq:E_condition_appendix} is automatically
satisfied by the solution for a nonzero asymptotic mass. The interface states therefore remain localized throughout
the whole momentum range in which the linear continuum description is applicable.

We next derive the general result near the projected Dirac point. At \(q=0\),
Eq.~\eqref{eq:matching_single_equation} becomes
\begin{equation}
\frac{M_L-E}{\sqrt{M_L^2-E^2}}
=
-
\frac{M_R-E}{\sqrt{M_R^2-E^2}} .
\label{eq:match_q_0}
\end{equation}
Squaring both sides gives
\begin{equation}
\frac{M_L-E}{M_L+E}
=
\frac{M_R-E}{M_R+E},
\end{equation}
and hence
\begin{equation}
2E(M_L-M_R)=0 .
\end{equation}
For \(M_L=M_R\), Eq.~\eqref{eq:match_q_0} has no solution inside the bulk
gap. A nontrivial bound state at \(q=0\) therefore has \(E=0\), and
substitution into Eq.~\eqref{eq:match_q_0} gives
\begin{equation}
\operatorname{sgn}(M_L)
=
-\operatorname{sgn}(M_R),
\label{eq:mass_sign_change_condition_appendix}
\end{equation}
which is the Jackiw--Rebbi normalizability condition~\cite{Jackiw_1976}.

For general interface parameters satisfying
Eq.~\eqref{eq:mass_sign_change_condition_appendix}, expansion of
Eq.~\eqref{eq:matching_single_equation} to first order in \(q\) gives
\begin{equation}
E(k)
=
\frac{|M_L||M_R|}{|M_L|+|M_R|}
\left[
\frac{u_y^R}{M_R}
-
\frac{u_y^L}{M_L}
\right]q
+
O(q^2).
\label{eq:general_slope_appendix}
\end{equation}
This expression may be written in terms of the valley Chern numbers of the two
bulk regions. 
For a two-level Hamiltonian $h_\alpha
=
\mathbf d_\alpha(p_x,q)\cdot\boldsymbol\sigma$, the Berry curvature is
\begin{equation*}
\Omega_\alpha(p_x,q)
=
-\frac{1}{2}
\frac{
\mathbf d_\alpha\cdot
\left(
\partial_{p_x}\mathbf d_\alpha
\times
\partial_q\mathbf d_\alpha
\right)
}{
|\mathbf d_\alpha|^3
}
\end{equation*}
so that from Eq.~\eqref{eq:H_side_alpha_appendix} we get
\begin{equation*}
\Omega_\alpha(p_x,q)= -
\frac{
M_\alpha v_\perp^\alpha u_y^\alpha
}{2
\left[
\left(v_\perp^\alpha p_x+u_x^\alpha q\right)^2
+
\left(u_y^\alpha q\right)^2
+
M_\alpha^2
\right]^{3/2}
}\,.
\end{equation*}
The valley Chern number is the integral of this Berry curvature over the
continuum momentum plane
\begin{equation}
C_{v,\alpha}
=
\frac{1}{2\pi}
\int\int 
dp_x\,dq\,
\Omega_\alpha(p_x,q).
\end{equation}
Performing the integral gives
\begin{equation}
C_{v,\alpha}
=
-\frac{1}{2}
\operatorname{sgn}
\left(
u_y^\alpha M_\alpha
\right),
\label{eq:valley_chern_alpha}
\end{equation}
where we used $v_\perp^\alpha>0$.
Therefore Eq.~\eqref{eq:general_slope_appendix} can be rewritten as
\begin{equation}
E(k)
=
2
\frac{|M_L||M_R|}{|M_L|+|M_R|}
\left[
C_{v,L}\frac{|u_y^L|}{|M_L|}
-
C_{v,R}\frac{|u_y^R|}{|M_R|}
\right]q .
\label{eq:general_slope_chern_appendix}
\end{equation}

\section{Characteristic Decay Length and Amplitude}
\label{app:Exponential_decay}

The penetration depth is set by the inverse real part of the decay exponent,
\begin{equation}
\xi_L=\frac{1}{\Re(\kappa_L)},
\qquad
\xi_R=\frac{1}{|\Re(\kappa_R)|}.
\end{equation}
Using Eqs.~\eqref{eq:kappa_L_decay} and \eqref{eq:kappa_R_decay}, one obtains
\begin{equation}
\xi_\alpha
=
\frac{|v_\perp^\alpha|}
{
\sqrt{
M_\alpha^2+\left(u_y^\alpha q\right)^2-E^2
}
}.
\label{eq:xi_alpha_general}
\end{equation}

Define the local bulk half-gap at fixed interface momentum as
\begin{equation}
\Delta_\alpha(k)
=
\sqrt{
M_\alpha^2+\left(u_y^\alpha q\right)^2
}.
\label{eq:bulk_half_gap_alpha_k}
\end{equation}
Then
\begin{equation}
\xi_\alpha(k,E)
=
\frac{|v_\perp^\alpha|}
{
\sqrt{
\Delta_\alpha(k)^2-E^2
}
}.
\label{eq:xi_gap_form}
\end{equation}
The full bulk band gap is $2\Delta_\alpha(k)$. Thus the penetration depth
diverges as $|E|\to\Delta_\alpha(k)$ and becomes shorter as the local bulk
gap increases.

Using Eq.~\eqref{eq:general_slope_chern_appendix}, to first order in $q$ this becomes

\begin{equation}
\xi_\alpha(k)
=
\frac{v_\perp^\alpha}{|M_\alpha|}
+
O(q^2)\,.
\label{eq:xi_alpha_small_q_expansion}
\end{equation}

Thus the interface eigenfunction may be written, to first order in \(q\), as
\begin{align}
\psi_L(x,k)
&=
A_L\,
\exp\!\left[
\frac{|M_L|}{v_\perp^L}\,x
\right]
\notag\\
&\hspace{1.2cm}\times
\exp\!\left[
-i\frac{u_x^L}{v_\perp^L}\,q x
\right]
\widehat\chi_L
+
O(q^2)\,,
\label{eq:interface_eigenfunction_transparent_explicit_L}
\\
\psi_R(x,k)
&=
A_R\,
\exp\!\left[
-\frac{|M_R|}{v_\perp^R}\,x
\right]
\notag\\
&\hspace{1.2cm}\times
\exp\!\left[
-i\frac{u_x^R}{v_\perp^R}\,q x
\right]
\widehat\chi_R
+
O(q^2)\,.
\label{eq:interface_eigenfunction_transparent_explicit_R}
\end{align}
The normalized eigenspinors are chosen as
\begin{equation}
\widehat\chi_L
=
\frac{\chi_L}
{\sqrt{\chi_L^\dagger\chi_L}},
\qquad
\widehat\chi_R
=
-
\frac{\chi_R}
{\sqrt{\chi_R^\dagger\chi_R}},
\qquad
\widehat\chi_\alpha^\dagger\widehat\chi_\alpha=1 .
\label{eq:normalized_chi_alpha}
\end{equation}
The minus sign in the definition of \(\widehat\chi_R\) is a convention adapted
to the transparent bound-state branch near \(q=0\). Indeed, after imposing the
matching condition~\eqref{eq:matching_single_equation}, near $q=0$ the two spinors may be written as
\begin{gather}
\chi_L
=
|M_L|
\begin{pmatrix}
-i\\
-r
\end{pmatrix},
\qquad
\chi_R
=
-|M_R|
\begin{pmatrix}
-i\\
-r
\end{pmatrix}
\label{eq:spinor_collinearity_near_q0}
\\
r\equiv
\frac{M_L-E}{|M_L|}
=
\frac{M_R-E}{-|M_R|}.
\notag
\end{gather}

Hence the two normalized spinors differ by a minus sign
with the original convention for \(\chi_\alpha\), and the choice in
Eq.~\eqref{eq:normalized_chi_alpha} gives
\begin{equation}
\widehat\chi_L=\widehat\chi_R
\end{equation}
on the interface branch near \(q=0\). The transparent boundary condition~\eqref{eq:transparent_matching_rotated_appendix}
therefore reduces to
\begin{equation}
A_L
=
\sqrt{
\frac{v_\perp^R}{v_\perp^L}
}\,
A_R
+
O(q^2).
\label{eq:A_matching_small_q}
\end{equation}

The remaining normalization is fixed by imposing
\begin{equation}
1
=
\int_{-\infty}^{0} dx\,
\psi_L^\dagger(x,k)\psi_L(x,k)
+
\int_{0}^{\infty} dx\,
\psi_R^\dagger(x,k)\psi_R(x,k) .
\label{eq:interface_mode_normalization_condition}
\end{equation}
One obtains
\begin{equation}
1
=
|A_L|^2
\frac{v_\perp^L}{2|M_L|}
+
|A_R|^2
\frac{v_\perp^R}{2|M_R|}
+
O(q^2).
\label{eq:A_normalization_condition_small_q}
\end{equation}
Together with Eq.~\eqref{eq:A_matching_small_q}, this gives
\begin{gather}
A_L
=
\sqrt{
\frac{
2|M_L||M_R|
}{
v_\perp^L\left(|M_L|+|M_R|\right)
}
}
+
O(q^2),
\label{eq:A_L_small_q}
\\
A_R
=
\sqrt{
\frac{
2|M_L||M_R|
}{
v_\perp^R\left(|M_L|+|M_R|\right)
}
}
+
O(q^2),
\label{eq:A_R_small_q}
\end{gather}
where the common global phase of the full eigenfunction has been fixed by choosing
\(A_L\) and \(A_R\) real and positive.

\section{Weighted Dispersion Relation}
\label{app:weighted_dispersion_relation}

Let \(P_L\) and \(P_R\) denote the normalized probability weights of the
interface state on the left and right sides,
\begin{equation}
P_L
=
\int_{-\infty}^{0} dx\,
\psi_L^\dagger\psi_L,
\qquad
P_R
=
\int_{0}^{\infty} dx\,
\psi_R^\dagger\psi_R .
\end{equation}
Using Eq.~\eqref{eq:interface_mode_normalization_condition} together with the amplitudes in
Eqs.~\eqref{eq:A_L_small_q} and \eqref{eq:A_R_small_q}, one obtains
\begin{align}
P_L
&=
\frac{|M_R|}
{|M_L|+|M_R|},
\notag\\
P_R
&=
\frac{|M_L|}
{|M_L|+|M_R|}.
\label{eq:transparent_left_right_weights}
\end{align}
These weights follow specifically from the transparent matching condition.
For more general matching conditions that preserve the common \(q=0\) spinor,
the envelope need not have the amplitudes \(A_L\) and \(A_R\) derived above.
In that case, \(P_L\) and \(P_R\) below denote the actual probability weights
and need not take the values in
Eq.~\eqref{eq:transparent_left_right_weights}.
Let \(\psi_0\) denote the normalized \(q=0\) interface state. By first-order
perturbation theory, the linear-in-\(q\) part of
Eq.~\eqref{eq:H_side_alpha_appendix} gives
\begin{equation}
E(k)
=
q
\left\langle
\psi_0
\left|
u_x\sigma_x+u_y\sigma_y
\right|
\psi_0
\right\rangle ,
\label{eq:first_order_linear_projection}
\end{equation}
where \(u_x\) and \(u_y\) are piecewise constant, equal to
\(u_x^\alpha\) and \(u_y^\alpha\) on side \(\alpha=L,R\).
Writing the wave function on each side as
\(\psi_{0,\alpha}(x)=f_\alpha(x)\widehat\chi\), the expectation value
factorizes as
\begin{align}
E(k)
&=
q\int_{-\infty}^{0}dx\,
\psi_{0,L}^\dagger
\left(
u_x^L\sigma_x+u_y^L\sigma_y
\right)
\psi_{0,L}
\notag\\
&\quad+
q\int_{0}^{\infty}dx\,
\psi_{0,R}^\dagger
\left(
u_x^R\sigma_x+u_y^R\sigma_y
\right)
\psi_{0,R}
\notag\\
&=
qP_L\,
\widehat\chi^\dagger
\left(
u_x^L\sigma_x+u_y^L\sigma_y
\right)
\widehat\chi\\
&\quad+
qP_R\,
\widehat\chi^\dagger
\left(
u_x^R\sigma_x+u_y^R\sigma_y
\right)
\widehat\chi
\notag\\
&=
\operatorname{sgn}(M_R)
\left(
u_y^L P_L+u_y^R P_R
\right)q .
\label{eq:weighted_slope_velocity}
\end{align}
In the last step, we used the fact that, from
Eq.~\eqref{eq:spinor_collinearity_near_q0}, the common normalized spinor is an
eigenstate of \(\sigma_y\), with
\begin{equation}
\widehat\chi^\dagger\sigma_y\widehat\chi
=
\operatorname{sgn}(M_R),
\qquad
\widehat\chi^\dagger\sigma_x\widehat\chi
=
0 .
\end{equation}

Equivalently, in terms of the valley Chern numbers, the expectation value can be written as
\begin{equation}
E(k)
=
2
\left(
C_{v,L}|u_y^L|P_L
-
C_{v,R}|u_y^R|P_R
\right)q .
\label{eq:weighted_slope_general_appendix}
\end{equation}

\section{Small momentum shifts and chemical-potential offsets}
\label{app:momentum_and_chemical_shifts}

We now allow the Dirac crossing and the local chemical potential to differ
slightly on the two sides of the interface. If the Dirac point on side
\(\alpha=L,R\) is shifted to \(K+\delta K_\alpha\), then the momentum measured
from the local cone is
\begin{equation}
q_\alpha
=
k-(K+\delta K_\alpha)
=
q-\delta K_\alpha .
\end{equation}
A common shift of the two Dirac points can be absorbed into the definition of
\(K\), while a common chemical-potential offset can be absorbed into the
energy origin. We therefore define the relative offsets
\begin{equation}
\delta K
=
\delta K_R-\delta K_L,
\qquad
\delta\mu
=
\delta\mu_R-\delta\mu_L,
\end{equation}
and adopt the symmetric convention
\begin{gather}
\delta K_L=-\frac{\delta K}{2},
\qquad
\delta K_R=\frac{\delta K}{2},
\\
\delta\mu_L=-\frac{\delta\mu}{2},
\qquad
\delta\mu_R=\frac{\delta\mu}{2}.
\label{eq:symmetric_shift_convention}
\end{gather}
The momenta measured from the two local cones are consequently
\begin{equation}
q_L=q+\frac{\delta K}{2},
\qquad
q_R=q-\frac{\delta K}{2}.
\end{equation}

In the locally rotated basis, the corresponding first-order perturbations are
\begin{align}
\delta h_L
&=
\frac{\delta K}{2}
\left(
u_x^L\sigma_x+u_y^L\sigma_y
\right)
+
\frac{\delta\mu}{2}\sigma_0,
\notag\\
\delta h_R
&=
-\frac{\delta K}{2}
\left(
u_x^R\sigma_x+u_y^R\sigma_y
\right)
-
\frac{\delta\mu}{2}\sigma_0 .
\label{eq:relative_shift_perturbation}
\end{align}
Projecting these perturbations onto the unperturbed interface mode gives
\begin{align}
\delta E
&=
\frac{\delta K}{2}\operatorname{sgn}(M_R)
\left(
u_y^L P_L-u_y^R P_R
\right)
+
\frac{\delta\mu}{2}
\left(
P_L-P_R
\right)
\notag\\
&=
\left(
C_{v,L}|u_y^L|P_L
+
C_{v,R}|u_y^R|P_R
\right)\delta K
+
\frac{P_L-P_R}{2}\,\delta\mu .
\label{eq:relative_shift_energy_correction}
\end{align}
The transparent-interface result follows by substituting the
weights in Eq.~\eqref{eq:transparent_left_right_weights}.

To first order, the correction is independent of \(q\).
The relative momentum and chemical-potential offsets therefore produce only
an energy offset in the interface dispersion. Possible velocity corrections
involving mixed terms such as \(q\,\delta K\) or \(q\,\delta\mu\) lie beyond
this joint first-order expansion.

\section{Second-order corrections in the kinetic channels}
\label{app:second_order_kinetic_corrections}

We now ask which second-order kinetic-channel terms give the leading
curvature corrections to the projected interface dispersion. The
second-order terms are treated perturbatively, and we retain only
contributions that are first order in their coefficients and at most first
order in the interface momentum \(q=k-K\). We therefore ignore terms
quadratic in the second-order coefficients, as well as terms contributing
only at \(O(q^2)\).
We now impose Hermiticity of the second-order operators across the interface by using the
minimal anticommutator prescription as in Eq.~\eqref{eq:sym_vxpx}. No additional independent interface
operator is introduced. Thus any
interface contribution which appears below is fixed by the Hermitian
ordering of the step-profile coefficients and is not an additional seam
parameter. As discussed below, a finite-width regularization is nevertheless required if a second-order coefficient is discontinuous or if the transparent matching condition produces a discontinuous unperturbed wave function.

With this convention, the interface dispersion is written as
\begin{equation*}
E(q)
=
\delta E_{0}
+
\left[
E_1^{(0)}
+
\delta E_{1}
\right]q
+
O(q^2,c^2)\,,
\end{equation*}
where \(E_1^{(0)}\) is the unperturbed linear coefficient derived in
Appendix~\ref{app:dispersion_near_q}, and \(c\) collectively denotes the quadratic kinetic
coefficients introduced through the
second-order expansion.

At \(q=0\), the Hamiltonian on side \(\alpha=L,R\) is
\begin{equation*}
h_{0,\alpha}
=
-i\,v_\perp^\alpha\sigma_x\partial_x
+
M_\alpha\sigma_z .
\end{equation*}
This unperturbed Hamiltonian has the chiral symmetry $\Gamma$ mapping an eigenstate of energy \(E\) to one of energy \(-E\)
\begin{gather*}
\Gamma=\sigma_y,
\qquad
\Gamma^2=1 ,\\
\Gamma h_0\Gamma^{-1}=-h_0 \,,
\end{gather*}
and a zero-energy mode can be chosen to have definite chirality
\begin{equation*}
\Gamma \psi_0=\eta\psi_0,
\qquad
\eta=\pm1 .
\end{equation*}
Furthermore, the Hamiltonian has the antiunitary symmetry $\Theta_{\mathrm r}$, 
\begin{gather*}
\Theta_{\mathrm r}=\sigma_z K,
\qquad
\Theta_{\mathrm r}^2=1 ,\\
\Theta_{\mathrm r}h_0\Theta_{\mathrm r}^{-1}=h_0 \,,
\end{gather*}
where \(K\) denotes complex conjugation. Therefore in analogy with a pure complex conjugation, the eigenstates of \(h_0\) may be chosen in a \(\Theta_{\mathrm r}\)-real basis. In particular, the zero mode may be chosen such that
\begin{equation*}
    \Theta_\mathrm{r}\psi_0=\psi_0\,.
\end{equation*}

We now allow second-order momentum corrections in the kinetic Pauli
channels already present in the rotated Dirac operator, namely
\(\sigma_x\) and \(\sigma_y\). The Hermitian
second-order expansion is
\begin{equation}
\delta h^{(2)}
=
\sum_{\mu=x,y}
\left[
\mathcal O_{xx}^{(\mu)}
+
q
\mathcal O_{xy}^{(\mu)}
+
\mathcal O_{yy}^{(\mu)} q^2
\right].
\label{eq:second_order_hermitian}
\end{equation}
where
\begin{align*}
\mathcal O_{xx}^{(\mu)}
&=
\frac12
\left\{
c_{xx}^{(\mu)}(x)\sigma_\mu,\hat p_x^2
\right\},
\\
\mathcal O_{xy}^{(\mu)}
&=
\frac12
\left\{
c_{xy}^{(\mu)}(x)\sigma_\mu,\hat p_x
\right\},
\\
\mathcal O_{yy}^{(\mu)}
&=
c_{yy}^{(\mu)}(x)\sigma_\mu \,.
\end{align*}
Here, the lower indices $i,j=x,y$ label real-space momentum directions,
whereas the parenthesized superscript $(\mu)$, with $\mu=x,y$, labels the
Pauli channel $\sigma_\mu$ in the locally rotated pseudospin frame.
The corresponding coefficients may change
across the interface,
\begin{equation*}
c_{ij}^{(\mu)}(x)
=
c_{ij}^{(\mu),L}\Theta(-x)
+
c_{ij}^{(\mu),R}\Theta(x),
\end{equation*}
but are taken to be constant inside each homogeneous region.
These anticommutators are the only
interface terms included in this minimal prescription.
Then the six kinetic-channel structures classified using the two symmetries are
\[
\begin{array}{@{}lcc@{}}
\hline
\text{Term} & \Gamma & \Theta_{\mathrm r} \\
\hline
\mathcal O_{xx}^{(x)}      & - & - \\
\mathcal O_{xy}^{(x)}     & - & + \\
\mathcal O_{yy}^{(x)}   & - & - \\
\mathcal O_{xx}^{(y)}      & + & + \\
\mathcal O_{xy}^{(y)}     & + & - \\
\mathcal O_{yy}^{(y)}   & + & + \\
\hline
\end{array}
\]

For a perturbation \(W(q)=q^mO\), the direct energy correction is
\begin{equation*}
\delta E_{\mathrm{dir}}
=
q^m\langle\psi_0|O|\psi_0\rangle .
\end{equation*}
Because we retain only terms through \(O(q)\), direct contributions with
\(m>1\) do not affect the projected dispersion at the order of interest, and we may neglect the terms $\mathcal{O}_{yy}^{(\mu)}$. Furthermore, for a chiral-odd operator satisfying $\Gamma \mathcal{O}\Gamma^{-1}=-\mathcal{O}$, the diagonal matrix elements on the chiral zero mode vanish
\begin{align*}
    \langle\psi_0|\mathcal O|\psi_0\rangle
&=
\langle \Gamma\psi_0|\Gamma\mathcal O \Gamma^{-1}|\Gamma\psi_0\rangle
\\
&=
-\langle\psi_0|\mathcal O|\psi_0\rangle
=
0\,.
\end{align*}

Therefore \(\mathcal O_{xx}^{(x)}\) does not contribute directly to
\(\delta E_0\), while \(q\mathcal O_{xy}^{(x)}\) does not contribute
directly to \(\delta E_1\).

Now consider the chiral-even operator $\mathcal O_{xx}^{(y)}$. For the
$q=0$ localized zero mode, write
\begin{equation*}
\psi_0(x)=f(x)\chi_\eta,
\qquad
\Gamma\chi_\eta=\eta\chi_\eta,
\end{equation*}
with $f(x)$ chosen real. For $M_LM_R<0$, the chirality is fixed by
\begin{equation}
\eta
=
-\operatorname{sgn}(M_L)
=
\operatorname{sgn}(M_R).
\label{eq:zero_mode_chirality_second_order}
\end{equation}
The contribution of $\mathcal O_{xx}^{(y)}$ depends on how the interface
parameters vary across the domain wall. The general result for a smooth
finite-width interface is derived in
Eq.~\eqref{app:smooth_delta_E0_integral}. Here, we consider the special case
in which the normal velocity and the quadratic coefficient remain spatially
constant,
\begin{equation*}
v_\perp(x)=v_\perp,
\qquad
c_{xx}^{(y)}(x)=c_{xx}^{(y)},
\end{equation*}
while the mass is smoothly interpolated according to
\begin{equation}
M(x)
=
\frac{M_L+M_R}{2}
+
\frac{M_R-M_L}{2}
\tanh\left(\frac{x}{w}\right),
\label{eq:smooth_tanh_mass_profile}
\end{equation}
with \(M_LM_R<0\).
For constant $v_\perp$, the smooth zero-mode equation in
Eq.~\eqref{eq:smooth_envelope_differential} reduces to
\begin{equation}
\partial_x f_w(x)
=
-\frac{\eta M(x)}{v_\perp}f_w(x).
\label{eq:tanh_mass_zero_mode_differential}
\end{equation}
The corresponding normalized envelope is
\begin{equation}
f_w(x)
=
\mathcal N_w
\exp\left[
-\frac{\eta(M_L+M_R)}{2v_\perp}x
\right]
\left[
\cosh\left(\frac{x}{w}\right)
\right]^{
-\frac{\eta(M_R-M_L)w}{2v_\perp}
}.
\label{eq:tanh_mass_zero_mode}
\end{equation}
Since $c_{xx}^{(y)}$ is constant, the second term in
Eq.~\eqref{app:smooth_delta_E0_integral} vanishes. The first-order energy
correction is therefore
\begin{align}
\delta E_0(w)
&=
\eta c_{xx}^{(y)}
\int_{-\infty}^{\infty}dx\,
\left|\partial_x f_w(x)\right|^2
\notag\\
&=
\frac{\eta c_{xx}^{(y)}}{v_\perp^2}
\int_{-\infty}^{\infty}dx\,
M(x)^2|f_w(x)|^2,
\label{eq:smooth_delta_E0_constant_parameters}
\end{align}
where the second equality follows directly from taking the modulus squared of Eq.~\eqref{eq:tanh_mass_zero_mode_differential}.
For the hyperbolic-tangent mass profile in
Eq.~\eqref{eq:smooth_tanh_mass_profile}, the integral can be evaluated
exactly and gives
\begin{equation}
\delta E_0(w)
=
\eta c_{xx}^{(y)}
\frac{|M_L||M_R|}
{v_\perp^2
\left[
1+\dfrac{\left(|M_L|+|M_R|\right)w}{v_\perp}
\right]}.
\label{eq:smooth_delta_E0_tanh_exact}
\end{equation}

Consequently, the zero-width limit is
\begin{align}
\lim_{w\rightarrow0}\delta E_0(w)
&=
\eta c_{xx}^{(y)}
\frac{|M_L||M_R|}{v_\perp^2}
\notag\\
&=
\eta c_{xx}^{(y)}
\left(
\frac{P_L}{\xi_L^2}
+
\frac{P_R}{\xi_R^2}
\right),
\label{eq:smooth_delta_E0_sharp_limit}
\end{align}
where the second equality follows from the sharp-interface weights in
Eq.~\eqref{eq:transparent_weights} and the decay lengths in
Eq.~\eqref{eq:xi_alpha_small_q}. Using
Eq.~\eqref{eq:zero_mode_chirality_second_order}, this may equivalently be
written as
\begin{equation}
\lim_{w\rightarrow0}\delta E_0(w)
=
\operatorname{sgn}(M_R)c_{xx}^{(y)}
\left(
\frac{P_L}{\xi_L^2}
+
\frac{P_R}{\xi_R^2}
\right).
\label{eq:smooth_delta_E0_sharp_limit_mass_sign}
\end{equation}
Thus, for equal normal velocities and a spatially constant
$c_{xx}^{(y)}$, a closed expression for the sharp-interface limit is
obtained directly from a smooth mass profile as $w\rightarrow0$. At every
finite $w$, the envelope and its derivatives are smooth. In the zero-width
limit, the envelope approaches the continuous piecewise-exponential
sharp-interface solution, while its first derivative develops a finite jump.
If either $v_\perp(x)$ or $c_{xx}^{(y)}(x)$ varies across the interface, the
correction must instead be evaluated from the general smooth-interface
expression in Eq.~\eqref{app:smooth_delta_E0_integral}. Its value then
depends on the detailed finite-width profiles of the interface parameters.

Finally, consider the direct contribution from
\(q\mathcal O_{xy}^{(y)}\). Again using
\(\psi_0=f\chi_\eta\), with \(f\) real,
\begin{align*}
q\langle\psi_0|\mathcal O_{xy}^{(y)}|\psi_0\rangle
&=
\frac{\eta q}{2}
\left[
\langle f|c_{xy}^{(y)}\hat p_x|f\rangle
+
\langle f|\hat p_x c_{xy}^{(y)}|f\rangle
\right]
\\
&=
-\frac{i\eta q}{2}
\int dx\,
\left[
f\,c_{xy}^{(y)}\,\partial_x f
+
f\,\partial_x
\left(c_{xy}^{(y)}f\right)
\right]
\\
&=
-\frac{i\eta q}{2}
\int dx\,
\partial_x
\left[
c_{xy}^{(y)}(x)|f(x)|^2
\right].
\end{align*}
Since \(f(x)\) is localized,
\begin{equation*}
\left.
c_{xy}^{(y)}(x)|f(x)|^2
\right|_{-\infty}^{+\infty}
=0 .
\end{equation*}
Therefore
\begin{equation*}
q\langle\psi_0|\mathcal O_{xy}^{(y)}|\psi_0\rangle
=0 .
\end{equation*}
Hence \(q\mathcal O_{xy}^{(y)}\) does not contribute directly to
\(\delta E_1\).

The leading correction to the coefficient of \(q\) can also arise from a
mixed second-order process between a perturbation \(q^m O\) and the linear
Dirac perturbation
\begin{equation*}
qV^{(1)}
=
q\left(u_x\sigma_x+u_y\sigma_y\right).
\end{equation*}
It is given by
\begin{equation*}
\delta E_{\mathrm{mix}}^{(2)}
=
2q^{m+1}\operatorname{Re}
\sum_{n\neq0}
\frac{
\langle\psi_0|O|\psi_n\rangle
\langle\psi_n|V^{(1)}|\psi_0\rangle
}{
-E_n
}.
\end{equation*}
This mixed term contributes to the coefficient of \(q\) only when \(m=0\).
Therefore, among the second-order kinetic structures, only
\(\mathcal O_{xx}^{(x)}\) and \(\mathcal O_{xx}^{(y)}\) can affect the
linear coefficient through such a mixed process.

The linear Dirac perturbations have the symmetry parities
\begin{gather*}
\Gamma(u_x\sigma_x)\Gamma^{-1}=-u_x\sigma_x,
\qquad
\Theta_{\mathrm r}(u_x\sigma_x)\Theta_{\mathrm r}^{-1}
=-u_x\sigma_x,
\\
\Gamma(u_y\sigma_y)\Gamma^{-1}=+u_y\sigma_y,
\qquad
\Theta_{\mathrm r}(u_y\sigma_y)\Theta_{\mathrm r}^{-1}
=+u_y\sigma_y .
\end{gather*}
Thus \(u_x\sigma_x\) is both chiral-odd and
\(\Theta_{\mathrm r}\)-odd, while \(u_y\sigma_y\) is both chiral-even and
\(\Theta_{\mathrm r}\)-even.

Because \(\Gamma h_0\Gamma^{-1}=-h_0\), the finite-energy states occur in chiral
pairs
\begin{equation*}
h_0(\Gamma\psi_n)=-E_n(\Gamma\psi_n).
\end{equation*}
Thus \(\psi_n\) and \(\Gamma\psi_n\) form a pair with energies \(E_n\) and
\(-E_n\).
For a product of two perturbations \(O\) and \(V\), with chiral parities
\(s_O\) and \(s_V\), the pair contribution is proportional to
\begin{gather*}
\frac{
\langle\psi_0|O|\psi_n\rangle
\langle\psi_n|V|\psi_0\rangle
}{
-E_n
}
+
\frac{
\langle\psi_0|O|\Gamma\psi_n\rangle
\langle\Gamma\psi_n|V|\psi_0\rangle
}{
+E_n
}\\
=
\left(s_Os_V-1\right)
\frac{
\langle\psi_0|O|\psi_n\rangle
\langle\psi_n|V|\psi_0\rangle
}{
E_n
}.
\end{gather*}
Therefore contributions with the same chiral parity cancel between the
\(\pm E_n\) partners. Only opposite-parity combinations survive. The only possible mixed terms are therefore
\(\mathcal O_{xx}^{(x)}\) with \(u_y\sigma_y\), and
\(\mathcal O_{xx}^{(y)}\) with \(u_x\sigma_x\).

For an operator \(A\) with definite \(\Theta_{\mathrm r}\)-parity
\begin{equation*}
\Theta_{\mathrm r}A\Theta_{\mathrm r}^{-1}
=
t_A A,
\qquad
t_A=\pm1 ,
\end{equation*}
antiunitarity gives
\begin{equation*}
\langle\psi_m|A|\psi_n\rangle^*
=
\langle\Theta_{\mathrm r}\psi_m|
\Theta_{\mathrm r}A\Theta_{\mathrm r}^{-1}
|\Theta_{\mathrm r}\psi_n\rangle
=
t_A
\langle\psi_m|A|\psi_n\rangle .
\end{equation*}
Therefore, if \(t_A=+1\), the matrix element is real, while if \(t_A=-1\),
the matrix element is purely imaginary.

The surviving chiral combinations always contain one
\(\Theta_{\mathrm r}\)-even operator and one \(\Theta_{\mathrm r}\)-odd
operator. Thus one of the two matrix elements is real and the other is purely
imaginary. Their product is therefore purely imaginary, so its real part
vanishes:
\begin{align*}
\operatorname{Re}
\left[
\langle\psi_0|\mathcal O_{xx}^{(x)}|\psi_n\rangle
\langle\psi_n|u_y\sigma_y|\psi_0\rangle
\right]
&=0,
\\
\operatorname{Re}
\left[
\langle\psi_0|\mathcal O_{xx}^{(y)}|\psi_n\rangle
\langle\psi_n|u_x\sigma_x|\psi_0\rangle
\right]
&=0 .
\end{align*} 
Consequently, the mixed second-order processes do not renormalize the linear
coefficient, and therefore
\begin{equation}
\delta E_1=0 .
\label{eq:delta_E1_zero_hermitian}
\end{equation}

The leading effects of the Hermitian second-order kinetic-channel
structures are summarized in Table~\ref{tab:second_order_summary}.

\begin{table}[t]
\centering
\scriptsize
\setlength{\tabcolsep}{2.4pt}
\caption{Leading effects of the Hermitian second-order kinetic-channel
structures.}
\label{tab:second_order_summary}
\begin{tabular}{@{}lcccc@{}}
\hline
Term & \(\delta E_0\) & \(\delta E_1\) \\
\hline
\(\mathcal O_{xx}^{(x)}\)      & \(0\) & \(0\) \\
\(\mathcal O_{xy}^{(x)}\)     & -- & \(0\) \\
\(\mathcal O_{yy}^{(x)}\)   & -- & -- \\
\(\mathcal O_{xx}^{(y)}\)     & $\delta E_{\mathrm{dir}}$ & \(0\) \\
\(\mathcal O_{xy}^{(y)}\)     & -- & \(0\) \\
\(\mathcal O_{yy}^{(y)}\)   & -- & -- \\
\hline
\end{tabular}
\end{table}

Combining these results, the single-cone interface dispersion through
linear order in \(q\) is
\begin{equation}
E(q)
=
\delta E_0
+
E_1^{(0)}q
+
O(q^2,c^2),
\label{eq:projected_dispersion_first_order_appendix}
\end{equation}
where \(\delta E_0\) comes from the $\mathcal O_{xx}^{(y)}$ term.
Thus, within the single-cone continuum theory with real coefficients,
transparent matching, and the minimal Hermitian anticommutator prescription,
the leading second-order kinetic-channel effect is to shift the interface
mode energy at \(q=0\). The linear slope on the other hand, is not modified.

It is however useful to distinguish the effect of these terms on the interface mode
from their effect on the homogeneous bulk bands. In a homogeneous region
\(\alpha\), the unperturbed Dirac Hamiltonian may be written as
\begin{equation*}
h_{0,\alpha}
=
\mathbf d_\alpha\cdot\boldsymbol\sigma ,
\qquad
\mathbf d_\alpha
=
\left(
v_\perp^\alpha q_x+u_x^\alpha q_y,\,
u_y^\alpha q_y,\,
M_\alpha
\right).
\end{equation*}
The quadratic kinetic-channel corrections considered above add only to the
\(\sigma_x\) and \(\sigma_y\) components,
\begin{equation*}
\delta h_\alpha^{(2)}
=
\delta d_x^\alpha\sigma_x
+
\delta d_y^\alpha\sigma_y 
=
\delta\mathbf d_\alpha \cdot \boldsymbol{\sigma},
\end{equation*}
with
\begin{align*}
\delta d_x^\alpha
&=
c_{xx}^{(x),\alpha}q_x^2
+
c_{xy}^{(x),\alpha}q_y q_x
+
c_{yy}^{(x),\alpha}q_y^2,
\\
\delta d_y^\alpha
&=
c_{xx}^{(y),\alpha}q_x^2
+
c_{xy}^{(y),\alpha}q_y q_x
+
c_{yy}^{(y),\alpha}q_y^2 .
\end{align*}
For the homogeneous bulk eigenstate \(|u_{\pm,\alpha}\rangle\), standard
first-order perturbation theory gives
\begin{equation*}
\delta E_{\pm,\alpha}^{(2)}
=
\langle u_{\pm,\alpha}|
\delta h_\alpha^{(2)}
|u_{\pm,\alpha}\rangle .
\end{equation*}
Since
\begin{equation*}
\langle u_{\pm,\alpha}|
\boldsymbol\sigma
|u_{\pm,\alpha}\rangle
=
\pm
\frac{\mathbf d_\alpha}{|\mathbf d_\alpha|},
\end{equation*}
this becomes
\begin{equation*}
\delta E_{\pm,\alpha}^{(2)}
=
\pm
\frac{
\mathbf d_\alpha\cdot\delta\mathbf d_\alpha
}{
|\mathbf d_\alpha|
}\,.
\end{equation*}
For a gapped bulk cone, \(M_\alpha\neq0\), one has
\begin{gather*}
|\mathbf d_\alpha|
=
|M_\alpha| = O(1),\\
d_x^\alpha,d_y^\alpha=O(|\mathbf{q}|),
\\
\delta d_x^\alpha,\delta d_y^\alpha=O(|\mathbf{q}|^2).
\end{gather*}
Hence
\begin{equation*}
\delta E_{\pm,\alpha}^{(2)}
=
O(k^3).
\end{equation*}
Thus the \(\sigma_x,\sigma_y\) quadratic kinetic-channel terms do not modify
the homogeneous gapped bulk dispersion through quadratic order in momentum.
They can nevertheless shift and renormalize the projected interface
dispersion.

\section{Smooth Finite-Width Interface}
\label{app:smooth_interface}

We now consider the case of a smooth finite-width interface.
We assume that the locally rotated Dirac parameters
\[
v_\perp(x),\qquad u_x(x),\qquad u_y(x),\qquad M(x)
\]
interpolate smoothly between their left and right asymptotic values. With the
same Hermitian ordering used in Eq.~\eqref{eq:sym_vxpx}, the zero-mode envelope at \(q=0\) is determined by  
\begin{equation}
\left[
v_\perp(x)\partial_x
+
\frac12\partial_x v_\perp(x)
+
\eta M(x)
\right]
f(x)
=
0 ,
\label{eq:smooth_envelope_differential}
\end{equation}
and therefore
\begin{equation}
f(x)
=
\frac{\mathcal N}{\sqrt{v_\perp(x)}}
\exp\left[
-\eta
\int_{0}^{x}
dx'\,
\frac{M(x')}{v_\perp(x')}
\right] .
\label{eq:smooth_envelope_solution}
\end{equation}
The normalization constant is fixed by
\begin{equation}
\int_{-\infty}^{\infty}dx\,|f(x)|^2=1 .
\label{eq:smooth_envelope_normalization}
\end{equation}
For an interface between asymptotic masses \(M_L\) and \(M_R\), a normalizable
mode requires \(M_LM_R<0\). The leading-order slope follows directly by projecting the \(V^{(1)}\) Dirac
perturbation onto the zero
mode:
\begin{align}
E_1^{(0)}
&=
\int_{-\infty}^{\infty}
dx\,
\psi_0^\dagger(x)
\left[
u_x(x)\sigma_x
+
u_y(x)\sigma_y
\right]
\psi_0(x)
\notag\\
&=
\int_{-\infty}^{\infty}
dx\,
|f(x)|^2
\chi_\eta^\dagger
\left[
u_x(x)\sigma_x
+
u_y(x)\sigma_y
\right]
\chi_\eta
\notag\\
&=
\int_{-\infty}^{\infty}
dx\,
|f(x)|^2
\left[
u_x(x)\chi_\eta^\dagger\sigma_x\chi_\eta
+
u_y(x)\chi_\eta^\dagger\sigma_y\chi_\eta
\right]
\notag\\
&=
\eta
\int_{-\infty}^{\infty}
dx\,
u_y(x)|f(x)|^2 .
\label{app:smooth_E1_zero}
\end{align}

The projected result in Eq.~\eqref{app:smooth_E1_zero} gives the generic
linear slope. As in the sharp-interface case, an additional reflection
symmetry can extend this cancellation to all orders in \(q\) within the
linear Dirac theory. To show this, the term \(q u_x(x)\sigma_x\) is first
removed by the phase transformation
\begin{equation}
\psi_q(x)
=
e^{-iq\vartheta(x)}
\widetilde\psi_q(x),
\qquad
\partial_x\vartheta(x)
=
\frac{u_x(x)}{v_\perp(x)} .
\label{eq:smooth_ux_gauge_transformation}
\end{equation}
Here \(\widetilde\psi_q(x)\) denotes the spinor wavefunction after the
position-dependent phase has been removed. Using the Hermitian ordering in Eq.~\eqref{eq:sym_vxpx}, the transformed
Hamiltonian is then
\begin{equation}
\widetilde h(q)
=
-\frac{i}{2}
\left\{
v_\perp(x)\sigma_x,
\partial_x
\right\}
+
q u_y(x)\sigma_y
+
M(x)\sigma_z .
\label{eq:smooth_gauge_transformed_hamiltonian}
\end{equation}
It is convenient to remove the position-dependent coefficient \(v_{\perp}(x)\) from the
normal kinetic operator by introducing the velocity-adapted coordinate~\cite{ghosh_2021}
\begin{equation}
s(x)
=
\int_0^x
\frac{dx'}{v_\perp(x')} .
\label{eq:smooth_velocity_adapted_coordinate}
\end{equation}
Under this change of coordinates, the normalized spinor wavefunction becomes
\begin{equation}
\Phi_q(s)
=
\sqrt{v_\perp(x(s))}\,
\widetilde\psi_q(x(s)).
\label{eq:smooth_rescaled_spinor}
\end{equation}
where the square-root factor preserves the normalization under
\(dx=v_\perp(x(s))\,ds\). This wavefunction rescaling cancels the derivative of \(v_\perp(x)\) generated by the
anticommutator in Eq.~\eqref{eq:smooth_gauge_transformed_hamiltonian}, so that
\begin{equation}
-\frac{i}{2}
\left\{
v_\perp(x)\sigma_x,
\partial_x
\right\}
\longrightarrow
-i\sigma_x\partial_s .
\end{equation}
The Hamiltonian therefore takes the simpler form
\begin{equation}
h_s(q)
=
-i\sigma_x\partial_s
+
q u_y(x(s))\sigma_y
+
M(x(s))\sigma_z .
\label{eq:smooth_adapted_hamiltonian}
\end{equation}
Now let \(\mathcal P_s\) denote the unitary reflection operator acting trivially
in pseudospin space. Its action on the rescaled two-component spinor
wavefunction is
\begin{equation}
\mathcal P_s\Phi_q(s)
=
\Phi_q(-s).
\end{equation}
It satisfies
\begin{equation}
\mathcal P_s^\dagger
=
\mathcal P_s^{-1}
=
\mathcal P_s,
\qquad
\mathcal P_s\partial_s\mathcal P_s^{-1}
=
-\partial_s,
\qquad
\mathcal P_s\boldsymbol{\sigma}\mathcal P_s^{-1}
=
\boldsymbol{\sigma}\,.
\end{equation}
Then, if the mass and tangential velocity are antisymmetric in the rescaled
coordinate,
\begin{equation}
M(x(-s))
=
-M(x(s)),
\qquad
u_y(x(-s))
=
-u_y(x(s)),
\label{eq:smooth_exact_flat_symmetry_appendix}
\end{equation}
then
\begin{equation}
\mathcal P_s
h_s(q)
\mathcal P_s^{-1}
=
-h_s(q).
\label{eq:smooth_reflection_anticommutation}
\end{equation}
The spectrum is therefore symmetric under \(E\rightarrow -E\) at each fixed
\(q\). 
At \(q=0\), the mass inversion produces the Jackiw--Rebbi zero mode derived
in Eqs.~\eqref{eq:smooth_envelope_differential} and
\eqref{eq:smooth_envelope_solution}. As long as the corresponding branch
remains an isolated nondegenerate interface eigenstate as \(q\) is varied,
it cannot move away from zero energy, since a nonzero eigenvalue would
require a distinct reflected partner at the opposite energy. Consequently, \(E(q)=0\)
throughout the whole momentum range over which the branch remains isolated and
nondegenerate. This is the smooth-interface counterpart of the exact
sharp-interface solution in
Eq.~\eqref{eq:exact_flat_dispersion_appendix}.

However, special care is required for a smooth reversal of \(u_y(x)\).
Such a profile necessarily passes through a point where the tangential
linear term \(q u_y(x)\sigma_y\) vanishes. Higher-order momentum terms or
additional bands may then become important in a microscopic realization.
The exact flatness above is therefore a property of the linear continuum
Hamiltonian, and its microscopic range of validity must be checked using a
lattice or multiband regularization.

The second-order kinetic-channel correction to the energy at \(q=0\) is
still controlled only by the chiral-even operator
\(\mathcal O_{xx}^{(y)}\), as discussed in
Appendix~\ref{app:second_order_kinetic_corrections}. For a smooth coefficient
\(c_{xx}^{(y)}(x)\), its contribution is the ordinary matrix element
\begin{equation}
\delta E_0
=
\left\langle
\psi_0
\left|
\frac12
\left\{
c_{xx}^{(y)}(x)\sigma_y,
\hat p_x^2
\right\}
\right|
\psi_0
\right\rangle .
\label{eq:smooth_delta_E0_matrix_element}
\end{equation}

Again, using \(\psi_0=f\chi_\eta\), with \(f\) real, this can be written as
\begin{align}
\delta E_0
&=
-\frac{\eta}{2}
\int dx\,
\left[
f\,c_{xx}^{(y)}\,\partial_x^2f
+
f\,\partial_x^2
\left(
c_{xx}^{(y)}f
\right)
\right]
\notag\\
&=
\eta
\int dx\,
c_{xx}^{(y)}(x)
|\partial_x f(x)|^2\\
&\quad
-
\frac{\eta}{2}
\int dx\,
\left[
\partial_x^2c_{xx}^{(y)}(x)
\right]
|f(x)|^2 .
\label{app:smooth_delta_E0_integral}
\end{align}
For smooth profiles with finite derivatives,
Eq.~\eqref{app:smooth_delta_E0_integral} is finite and unambiguous at every
nonzero interface width. The special case of constant $v_\perp$ and
$c_{xx}^{(y)}$, together with the hyperbolic-tangent mass profile in
Eq.~\eqref{eq:smooth_tanh_mass_profile}, is evaluated explicitly in
Appendix~\ref{app:second_order_kinetic_corrections}. Its zero-width limit is
given by Eq.~\eqref{eq:smooth_delta_E0_sharp_limit_mass_sign}.
Consequently, within the same perturbative and symmetry assumptions as in
Appendix~\ref{app:second_order_kinetic_corrections}, the smooth-interface projected
dispersion through linear order in \(q\) still has the same form as in Eq.~\eqref{eq:projected_dispersion_first_order_appendix}. The finite-width profile therefore changes the numerical values of
\(E_1^{(0)}\) and \(\delta E_0\) through the envelope \(f(x)\) and the smooth
coefficient profiles, but it does not introduce a new linear-in-\(q\)
second-order correction under the minimal Hermitian prescription.

\section{Two-cone continuum model}
\label{app:two_cone_continuum_model}

We now generalize the interface problem to two Dirac cones. The cone index is
described by Pauli matrices \(\tau_{1,2,3}\), and we define the projectors
\begin{equation*}
\begin{aligned}
P_1
&=
\frac{\tau_0+\tau_3}{2},
\\
P_2
&=
\frac{\tau_0-\tau_3}{2}.
\end{aligned}
\end{equation*}
For cone \(\beta=1,2\), let \(K_\beta\) denote the projected Dirac momentum
along the interface direction and define
\begin{equation}
q_\beta
\equiv
k-K_\beta .
\label{eq:q_beta_def}
\end{equation}

The following construction should be understood as a two-sector low-energy theory. It is appropriate when the two cones represent independent low-energy sectors, such as distinct microscopic bands or degrees of freedom. Since Dirac cones in lattice models occur in pairs, this setting is most naturally realized in interface problems for systems with more than one pair of Dirac cones, where one considers crossings between interface modes originating from cones that are not related by fermion doubling and therefore do not belong to the same globally connected band pair. It should not be interpreted as a full-Brillouin-zone description of two Dirac points of a single lattice band pair related by fermion doubling. In that case, the two valley continuum modes are only local Dirac-point limits of one globally connected lattice interface band, whose connection away from the cone projections is controlled by the full lattice Hamiltonian. Describing that interface band and any possible hybridization therefore requires solving the full lattice interface problem, typically numerically, and lies outside the predictive scope of this two-cone continuum model.

We work in the locally rotated basis of each cone, so that the normal kinetic
matrix is proportional to \(\sigma_x\). The globally Hermitian uncoupled
two-cone Hamiltonian is block diagonal in cone space,
\begin{equation}
H_0
=
\sum_{\beta=1}^2
P_\beta\otimes h_\beta ,
\label{eq:H_twocone_uncoupled}
\end{equation}
with
\begin{align}
h_\beta
&=
-i
\left[
v_{\perp,\beta}(x)\sigma_x\partial_x
+
\frac12
\left(
\partial_x v_{\perp,\beta}(x)
\right)
\sigma_x
\right]
\notag\\
&\quad
+
q_\beta
\left[
u_{x,\beta}(x)\sigma_x
+
u_{y,\beta}(x)\sigma_y
\right]
+
M_\beta(x)\sigma_z .
\label{eq:h_beta_global_rotated}
\end{align}
Before inter-cone coupling is introduced, this Hamiltonian is simply the
direct sum of two independent copies of the single-cone interface problem
derived above. Consequently, all existence conditions, matching conditions,
decay lengths, and projected dispersions apply cone by cone, with the
replacement
\[ \left( v_\perp, u_x, u_y, M, q \right) \to \left( v_{\perp,\beta}, u_{x,\beta}, u_{y,\beta}, M_\beta, q_\beta \right). \]

For a sharp interface at \(x=0\), the cone-resolved parameters are taken to be
piecewise constant,
\begin{equation}
X_\beta(x)
=
X_\beta^L\Theta(-x)
+
X_\beta^R\Theta(x),
\label{eq:twocone_step_profile}
\end{equation}
where
\begin{equation*}
X_\beta
\in
\left\{
v_{\perp,\beta},
u_{x,\beta},
u_{y,\beta},
M_\beta
\right\}.
\end{equation*}
Thus, away from the interface, on side \(\alpha=L,R\), the Hamiltonian for
cone \(\beta\) becomes
\begin{align}
h_{\beta,\alpha}
&=
-i
v_{\perp,\beta}^\alpha
\sigma_x\partial_x
\notag\\
&\quad
+
q_\beta
\left[
u_{x,\beta}^\alpha\sigma_x
+
u_{y,\beta}^\alpha\sigma_y
\right]
+
M_\beta^\alpha\sigma_z .
\label{eq:h_beta_side_alpha}
\end{align}
Equivalently,
\begin{equation}
H_0^\alpha
=
\sum_{\beta=1}^2
P_\beta\otimes h_{\beta,\alpha}.
\label{eq:H_twocone_side_alpha}
\end{equation}

For each cone separately, the transparent matching condition in the locally
rotated basis is
\begin{equation}
\psi_{\beta,L}(0)
=
\sqrt{
\frac{
v_{\perp,\beta}^R
}{
v_{\perp,\beta}^L
}
}
\,
\psi_{\beta,R}(0).
\label{eq:twocone_transparent_matching}
\end{equation}
A localized interface branch exists for cone \(\beta\) when the Dirac mass
changes sign across the interface,
\begin{equation}
M_\beta^L M_\beta^R <0 .
\label{eq:twocone_mass_sign_condition}
\end{equation}

When Eq.~\eqref{eq:twocone_mass_sign_condition} holds, cone \(\beta\)
contributes one low-energy interface branch. To linear order in
\(q_\beta\), and before including inter-cone coupling, its dispersion is
\begin{equation}
\varepsilon_\beta(k)
=
s_\beta q_\beta
+
O(q_\beta^2),
\label{eq:eps_beta_linear}
\end{equation}
where the cone-resolved interface velocity is the corresponding single-cone
result,
\begin{align}
s_\beta
&=
\frac{
|M_\beta^L||M_\beta^R|
}{
|M_\beta^L|+|M_\beta^R|
}
\notag\\
&\quad\times
\left[
\frac{
u_{y,\beta}^R
}{
M_\beta^R
}
-
\frac{
u_{y,\beta}^L
}{
M_\beta^L
}
\right].
\label{eq:s_beta_general}
\end{align}
Thus, in the absence of inter-cone coupling, the interface spectrum is just
the union of the independent single-cone spectra. 

If the second-order kinetic-channel corrections discussed in
Appendix~\ref{app:second_order_kinetic_corrections} are retained, then the projected
single-cone result again applies separately to each cone. In that case
Eq.~\eqref{eq:eps_beta_linear} is replaced by
\begin{equation}
\varepsilon_\beta(k)
=
\delta E_{0,\beta}
+
s_\beta q_\beta
+
O(q_\beta^2,c^2),
\label{eq:eps_beta_linear_with_shift}
\end{equation}
where
\begin{align}
\delta E_{0,\beta}
&=
\left\langle
\psi_{\beta,0}
\left|
\frac12
\left\{
c_{xx,\beta}^{(y)}(x)\sigma_y,
\hat p_x^2
\right\}
\right|
\psi_{\beta,0}
\right\rangle .
\label{eq:delta_E0_beta_def}
\end{align}
Here \(\psi_{\beta,0}\) is the normalized zero-mode wavefunction of cone
\(\beta\) at \(q_\beta=0\). As in the single-cone calculation,
\(\delta E_{0,\beta}\) depends on the microscopic or smoothing prescription
when \(c_{xx,\beta}^{(y)}(x)\) is discontinuous. The linear coefficient
\(s_\beta\) is not renormalized by these second-order kinetic-channel terms
within the minimal Hermitian prescription used above.

We next include coupling between the two interface bands. Such a coupling
is meaningful in the projected low-energy theory only when the two interface
modes belong to the same conserved-\(k\) sector. Since \(k\) is conserved
along a translationally invariant interface, inter-cone coupling is allowed
only when the two projected cones are momentum-compatible. That is,
\begin{equation}
K_1-K_2
=
G_\parallel ,
\label{eq:momentum_matching_condition}
\end{equation}
where \(G_\parallel\) is a reciprocal lattice vector of the interface, or the
interface perturbation itself must carry the missing longitudinal momentum.
The continuum coupling written below assumes that this momentum-matching
condition has been satisfied. Otherwise the projected coupling \(g(k)\)
vanishes for a clean translationally invariant interface.

The most general Hermitian inter-cone coupling has the form
\begin{align}
W
&=
\tau_+\otimes \Gamma_{12}(x)
+
\tau_-\otimes \Gamma_{12}^\dagger(x),
\label{eq:W_general_hermitian}
\\
\tau_\pm
&=
\frac{\tau_1\pm i\tau_2}{2}.
\notag
\end{align}
Here \(\Gamma_{12}(x)\) is a \(2\times2\) matrix acting in the Dirac spinor
space. If \(\Gamma_{12}=\Gamma_{12}^\dagger\equiv\Gamma\), this reduces to
\begin{equation}
W
=
\tau_1\otimes\Gamma(x).
\label{eq:W_tau_x_form}
\end{equation}

If the inter-cone coupling is first written in the original pseudospin basis
as \(\Gamma_{12}^{\mathrm{orig}}(x)\), then the coupling matrix appearing in
the locally rotated cone bases is
\begin{equation}
\Gamma_{12}(x)
=
S_1^\dagger(x)
\Gamma_{12}^{\mathrm{orig}}(x)
S_2(x).
\label{eq:Gamma12_rotated_basis}
\end{equation}
Thus even a scalar microscopic coupling can acquire a nontrivial spinor structure in the rotated basis when the two cones are rotated by different matrices. Since \(H_0\) is diagonal in the cone index, while the inter-cone operator \(W\) is off-diagonal, projecting \(H_0+W\) onto the two interface modes gives the effective edge Hamiltonian
\begin{equation}
H_{\mathrm{edge}}(k)
=
\begin{pmatrix}
\varepsilon_1(k) & g(k) \\
g^*(k) & \varepsilon_2(k)
\end{pmatrix},
\label{eq:H_edge_basic}
\end{equation}
where
\begin{align}
g(k)
&=
\int_{-\infty}^{\infty} dx\,
\psi_1^\dagger(x,k)\Gamma_{12}(x)\psi_2(x,k).
\label{eq:g_def_general}
\end{align}
We define the contact coupling as the limit of a narrow but finite-width profile
\begin{equation}
\Gamma_{12}(x)
=
\Gamma_0 f_w(x),
\label{eq:contact_regularization}
\end{equation}
where \(f_w(x)\) is localized near \(x=0\), has width \(w\), and is normalized as
\begin{equation}
\int_{-\infty}^{\infty} dx\, f_w(x) = 1 .
\label{eq:fw_normalization}
\end{equation}
For example, one may take \(f_w(x)\) to be a narrow Gaussian or any other smooth function that approaches \(\delta(x)\) as \(w\to0\). For finite \(w\), the coupling is then
\begin{equation}
g_w(k)
=
\int_{-\infty}^{\infty} dx\,
f_w(x)\psi_1^\dagger(x,k)\Gamma_0\psi_2(x,k).
\label{eq:gw_regularized_contact}
\end{equation}
With this regularization we can write the sharp-contact expression
\begin{equation}
g(k)
\to
\psi_1^\dagger(0,k)\Gamma_0\psi_2(0,k).
\label{eq:g_contact_formal}
\end{equation}
Thus Eq.~\eqref{eq:g_contact_formal} should be regarded as shorthand for a chosen microscopic or finite-width regularization, not as an independent universal sharp-interface prescription. The eigenvalues of the projected two-mode Hamiltonian are
\begin{align}
E_\pm(k)
&=
\frac{ \varepsilon_1(k)+\varepsilon_2(k) }{2}
\notag\\
&\quad
\pm
\sqrt{
\left[
\frac{ \varepsilon_1(k)-\varepsilon_2(k) }{2}
\right]^2
+
|g(k)|^2
}.
\label{eq:Epm_avoided_gk}
\end{align}
The uncoupled bands cross at momenta \(k=k_c\) satisfying
\begin{equation}
\varepsilon_1(k_c) = \varepsilon_2(k_c),
\label{eq:crossing_condition}
\end{equation}
provided such a solution lies within the regime where both cone expansions remain valid and both interface states remain inside the corresponding bulk gaps. If \(g(k_c)\neq0\), the crossing is replaced by an avoided crossing with gap
\begin{equation}
\Delta_{\mathrm{gap}}
=
2|g(k_c)|.
\label{eq:gap_2gk}
\end{equation}
If \(g(k_c)=0\), either because of symmetry or because longitudinal momentum conservation forbids inter-cone scattering, the crossing remains ungapped at this order. If the two uncoupled bands do not cross within the low-energy window, the coupling only hybridizes and repels the two interface modes without producing a local avoided crossing.

\section{Minimal Lattice Model}
\label{app:minimal_lattice_model}

We take a square Brillouin zone and place the two Dirac cones at the high
symmetry points
\begin{equation}
\Gamma=(0,0),
\qquad
M=(\pi,\pi).
\label{eq:lattice_cones_Gamma_M}
\end{equation}
In this section we consider an interface parallel to the \(y\) direction, so
that \(k_y\) is the conserved interface momentum. The two cones then project
to
\begin{equation}
K_\Gamma=0,
\qquad
K_M=\pi ,
\label{eq:projected_Gamma_M}
\end{equation}
which are separated by half of the one-dimensional interface Brillouin zone.
Thus a relative momentum shift \(\delta K=\pi\) along the interface exchanges
the two projected Dirac cones.

We first define the massless lattice model. The Bloch Hamiltonian is
\begin{equation}
H_0(\mathbf k)
=
d_x(\mathbf k)\sigma_x
+
d_y(\mathbf k)\sigma_y .
\label{eq:H0_lattice_bloch_pauli}
\end{equation}
Its band energies are
\begin{equation}
E_\pm(\mathbf k)
=
\pm
\sqrt{
d_x(\mathbf k)^2
+
d_y(\mathbf k)^2
}.
\label{eq:lattice_band_energies}
\end{equation}
Therefore, a gap closing requires
\begin{equation}
d_x(\mathbf k)=0,
\qquad
d_y(\mathbf k)=0,
\label{eq:dirac_zero_conditions}
\end{equation}
which occurs only at the Dirac points. Hence, two independent momentum-dependent functions are needed. If only
one Pauli channel were present, then the
condition \(d_x(\mathbf k)=0\) would generically define a nodal line in two
dimensions rather than isolated Dirac points. 

Since the target points \(\Gamma\) and \(M\) lie on the diagonal
\(k_x=k_y\), it is useful to introduce rotated momentum coordinates
\begin{equation}
\tilde k_x
=
\frac{k_x+k_y}{2},
\qquad
\tilde k_y
=
\frac{k_x-k_y}{2}.
\label{eq:rotated_lattice_momenta}
\end{equation}
The coordinate \(\tilde k_x\) parametrizes motion along the
\(\Gamma\)-\(M\) diagonal, while \(\tilde k_y\) parametrizes motion transverse
to it. In these coordinates,
\begin{equation}
\Gamma:
\quad
(\tilde k_x,\tilde k_y)=(0,0),
\end{equation}
while \(M\) is represented by boundary points such as
\begin{equation}
M:
\quad
(\tilde k_x,\tilde k_y)=(\pi,0),
\end{equation}
with equivalent representatives obtained by reciprocal-lattice translations.
A minimal function vanishing only when its arguments are \(0\) or \(\pm\pi\) is a sine function. Therefore the simultaneous conditions
\begin{equation}
d_x^{(0)}(\mathbf k)
=
\sin\tilde k_x,
\qquad
d_y^{(0)}(\mathbf k)
=
-\sin\tilde k_y .
\label{eq:unperiodic_dx_dy}
\end{equation}
select only \(\Gamma\) and the reciprocal-lattice-equivalent representatives
of \(M\) within the rotated square Brillouin zone.

This gives the desired band energies,
\begin{equation}
E_\pm^{(0)}(\mathbf k)
=
\pm
\sqrt{
\sin^2\tilde k_x
+
\sin^2\tilde k_y
},
\label{eq:unperiodic_band_energies}
\end{equation}
with zeros only at \(\Gamma\) and \(M\), up to reciprocal-lattice
equivalence. However, the Hamiltonian itself is not periodic in the original
square Brillouin zone. Under \(k_x\to k_x+2\pi\), one has
\begin{equation}
\tilde k_x\to\tilde k_x+\pi,
\qquad
\tilde k_y\to\tilde k_y+\pi,
\end{equation}
and therefore
\begin{equation}
\sin\tilde k_x\to-\sin\tilde k_x,
\qquad
\sin\tilde k_y\to-\sin\tilde k_y .
\end{equation}
Thus both \(d_x^{(0)}\) and \(d_y^{(0)}\) change sign. The spectrum is
periodic, but the Bloch Hamiltonian changes as
\begin{equation}
H_0(\mathbf k+\mathbf G)
=
-H_0(\mathbf k).
\end{equation}

To restore periodicity without changing the band energies, we multiply the
off-diagonal element by a phase factor of unit modulus
\begin{equation}
e^{i\tilde k_x}
\left[
\sin\tilde k_x
+
i\sin\tilde k_y
\right].
\label{eq:offdiagonal_D_lattice}
\end{equation}
The factor \(e^{i\tilde k_x}\) never vanishes and satisfies
\begin{equation}
\left|e^{i\tilde k_x}\right|^2=1 .
\end{equation}
It therefore does not change the band energies.
Its role is instead to restore the lattice periodicity, since under either primitive
reciprocal-lattice translation, both the bracket in
Eq.~\eqref{eq:offdiagonal_D_lattice} and
\(e^{i\tilde k_x}\) change sign. Thus the two sign changes cancel and
\begin{equation}
H_0^{(0)}(\mathbf k+\mathbf G)
=
H_0^{(0)}(\mathbf k).
\end{equation}

Eq.~\eqref{eq:offdiagonal_D_lattice} can be written as a finite
Fourier series:
\begin{align*}
e^{i\tilde k_x}
\left[
\sin\tilde k_x
+
i\sin\tilde k_y
\right]
&=
\frac{
e^{i(k_x+k_y)}-1
}{2i}
+
\frac{
e^{ik_x}-e^{ik_y}
}{2}
\notag\\
&=
\frac{i}{2}
+
\frac{1}{2}e^{ik_x}
-
\frac{1}{2}e^{ik_y}
-
\frac{i}{2}e^{i(k_x+k_y)} .
\label{eq:D_fourier_expanded}
\end{align*}
Therefore, the nonzero inter-sublattice
hoppings are
\begin{equation}
\label{eq:lattice_hoppings_appendix}
\begin{aligned}
t_{\mathbf 0}
&=
\frac{i}{2},
&
t_{\hat x}
&=
\frac{1}{2},
\\
t_{\hat y}
&=
-\frac{1}{2},
&
t_{\hat x+\hat y}
&=
-\frac{i}{2}.
\end{aligned}
\end{equation}

We can introduce an anisotropy parameter \(r\) in the massless off-diagonal element,
\begin{equation}
e^{i\tilde k_x}
\left[
\sin\tilde k_x
+
i r \sin\tilde k_y
\right],
\qquad r>0 .
\label{eq:anisotropic_D_lattice}
\end{equation}
The isotropic model discussed above corresponds to \(r=1\). The anisotropy changes the local Dirac
velocities, but it does not displace nor introduce additional cones.

The Fourier expansion of Eq.~\eqref{eq:anisotropic_D_lattice} is
\begin{equation}
\frac{i}{2}
+
\frac{r}{2}e^{ik_x}
-
\frac{r}{2}e^{ik_y}
-
\frac{i}{2}e^{i(k_x+k_y)} .
\label{eq:anisotropic_D_fourier}
\end{equation}

At a fixed relative momentum \(k_\alpha\), the off-diagonal matrix element of Eq.~\eqref{eq:anisotropic_D_fourier} is
\[D_r(k_x,k_\alpha) =
t_{\mathrm{intra}}(k_\alpha)
+
t_{\mathrm{inter}}(k_\alpha)e^{ik_x},
\]
with
\[
t_{\mathrm{intra}}(k_\alpha)
=
\frac{i}{2}
-
\frac{r}{2}e^{ik_\alpha},
\qquad
t_{\mathrm{inter}}(k_\alpha)
=
\frac{r}{2}
-
\frac{i}{2}e^{ik_\alpha}
\]
the \(k_\alpha\)-dependent intracell and intercell hoppings. As
\(k_x\) winds around the one-dimensional Brillouin zone,
\(D_r(k_x,k_\alpha)\) traces a circle in the complex plane centered at
\(t_{\mathrm{intra}}(k_\alpha)\) with radius
\(|t_{\mathrm{inter}}(k_\alpha)|\) so that the corresponding winding number is nonzero when this circle encloses the origin, namely when
\[
|t_{\mathrm{inter}}(k_\alpha)|
>
|t_{\mathrm{intra}}(k_\alpha)|.
\]
In our unit cell convention, this corresponds to the existence of an isolated edge state. Conversely,
\[
|t_{\mathrm{inter}}(k_\alpha)|
<
|t_{\mathrm{intra}}(k_\alpha)|
\]
corresponds to the case which does not support an isolated edge state. 

The criterion can be evaluated explicitly:
\[
|t_{\mathrm{inter}}(k_\alpha)|^2
=
\left|
\frac{r}{2}
-
\frac{i}{2}e^{ik_\alpha}
\right|^2
=
\frac{1+r^2+2r\sin k_\alpha}{4},
\]
whereas
\[
|t_{\mathrm{intra}}(k_\alpha)|^2
=
\left|
\frac{i}{2}
-
\frac{r}{2}e^{ik_\alpha}
\right|^2
=
\frac{1+r^2-2r\sin k_\alpha}{4}.
\]
Therefore, for \(r>0\), the edge-state condition for this termination is
\[
\sin k_\alpha>0.
\]

For the unshifted interface, \(\delta K=0\), the minimum of
\(|t_{\mathrm{int}}(k_y)|\) occurs at
\[
k_y=-\frac{\pi}{2}.
\]
Both domains have the same local momentum,
\[
k_L=k_R=-\frac{\pi}{2}
\implies
\sin k_L=\sin k_R=-1.
\]
Hence neither of the decoupled terminations support
isolated edge states and the
interface branch is therefore not protected from approaching the bulk continuum, even though the anisotropic hopping keeps the bare seam matrix element finite.

For the shifted interface, \(\delta K=\pi\), the minimum of
\(|t_{\mathrm{int}}(k_y)|\) occurs at
\[
k_y=0,
\]
with local momenta
\[
k_L=0,
\qquad
k_R=-\pi .
\]
In both domains,
\[
\sin k_\alpha=0.
\]
The corresponding decoupled half-space chains are therefore at a transition point, rather than in a fully
localized-edge-state regime.

\subsection{Wilson-regularized lattice model}
\label{app:wilson_edge_state_model}

The model uses the standard Wilson-Dirac regularization of a lattice Dirac
Hamiltonian. In this regularization a momentum-dependent mass is added so that
the low-energy Dirac point is left unchanged, while the lattice doubler is
assigned a different mass \cite{Kuno_2018}.
We consider
\[
H_{\mathrm W}(\mathbf k)
=
\sin k_x\,\sigma_x
+
v_y\sin k_y\,\sigma_y
+
\left[
m+B(1-\cos k_x)
\right]\sigma_z .
\]
For an interface parallel to \(y\), \(k_y\) is conserved. At each fixed
\(k_y\), the part that controls localization in the \(x\) direction is a
one-dimensional Wilson-regularized Dirac chain,
\[
H_x(k_x)
=
\sin k_x\,\sigma_x
+
\left[
m+B(1-\cos k_x)
\right]\sigma_z .
\]
This one-dimensional Hamiltonian has a chiral symmetry,
\[
\{\sigma_y,H_x(k_x)\}=0 .
\]
As is conventional for chiral two-band chains, one can rotate to the basis in
which the chiral operator is diagonal. In this basis \(H_x\) is purely
off-diagonal,
\[
H_x(k_x)
\sim
\begin{pmatrix}
0 & q_m(k_x) \\
q_m^*(k_x) & 0
\end{pmatrix},
\]
with
\[
q_m(k_x)
=
m+B(1-\cos k_x)
+
i\sin k_x .
\]
The winding of \(q_m(k_x)\)
around the origin determines whether the one-dimensional chain supports an
edge-state. As \(k_x\) winds around the Brillouin zone, the curve
\(q_m(k_x)\) encloses the origin for
\[
-2B<m<0 .
\]
Thus, in this range, the corresponding one-dimensional chain supports a
localized edge state for the chosen termination while for \(m>0\), the same termination is trivial. The remaining term
\(v_y\sin k_y\,\sigma_y\) disperses this state along the interface, but does
not change the winding of the \(x\)-direction chain.

The role of the momentum-dependent mass term can be seen directly from the
two zeros of the \(x\)-direction kinetic term. Without it, \(\sin k_x\) vanishes both at
\(k_x=0\) and at \(k_x=\pi\). The additional term \(B(1-\cos k_x)\sigma_z\) vanishes at \(k_x=0\),
but is finite at \(k_x=\pi\),
\[
B(1-\cos \pi)=2B .
\]
Thus the effective masses at the two Dirac cones are \(m\) and
\(m+2B\), respectively. Thus, for
\[
0<m_0<2B ,
\]
the domain wall
\[
m_L=-m_0,
\qquad
m_R=+m_0
\]
inverts the mass at \(k_x=0\), but leaves the doubled sector at \(k_x=\pi\)
non-inverted. The momentum-dependent mass term therefore both removes the extra
domain-wall mode from the lattice doubler and gives the fixed-\(k_y\) chain a
nonzero winding over the full interface Brillouin zone. Without it, the chiral off-diagonal block would be
\[
q_m(k_x)=m+i\sin k_x ,
\]
which only moves along a vertical line in the complex plane and therefore
does not wind around the origin.

\clearpage
\bibliographystyle{apsrev4-2}
\bibliography{bibliography-28-08-2024-11}

\end{document}